\documentclass[journal,comsoc,twocolumn]{IEEEtran}
\usepackage{amsthm,amsmath,amssymb}
\usepackage{mathrsfs}
\usepackage{amsfonts,amssymb}
\usepackage[T1]{fontenc}
\usepackage{hyperref}
\usepackage{multirow}
\usepackage{diagbox}
\usepackage{cite}
\usepackage{color}
\usepackage[pdftex]{graphicx}
\usepackage{graphicx, subfigure}
\usepackage{amsmath}
\usepackage[cmintegrals]{newtxmath}
\usepackage{threeparttable}
\begin{document}
\newcommand{\tabincell}[2]{\begin{tabular}{@{}#1@{}}#2\end{tabular}}

\title{Deep Learning-Based CSI Feedback for Beamforming in Single- and Multi-cell Massive MIMO Systems}

\author{\normalsize {Jiajia~Guo, 
Chao-Kai~Wen, \IEEEmembership{\normalsize {Member,~IEEE}},
~and  Shi~Jin, \IEEEmembership{\normalsize {Senior Member,~IEEE}}
}

\thanks{This work was supported in part by the National Key Research and Development Program 2018YFA0701602, the National Science Foundation of China (NSFC) for Distinguished Young Scholars with Grant 61625106, and the NSFC under Grant 61941104.
The work of C.-K. Wen was supported in part by the Ministry of Science and Technology of Taiwan
under grants MOST 108-2628-E-110-001-MY3. (Corresponding author: Shi Jin)
}
\thanks{J.~Guo and S.~Jin are with the National Mobile Communications Research
Laboratory, Southeast University, Nanjing, 210096, P. R. China (email: jiajiaguo@seu.edu.cn, jinshi@seu.edu.cn).}
\thanks{C.-K.~Wen is with the Institute of Communications Engineering, National Sun Yat-sen University, Kaohsiung 80424, Taiwan (e-mail: chaokai.wen@mail.nsysu.edu.tw).}
}

\maketitle

\begin{abstract}
The potentials of massive multiple-input multiple-output (MIMO) are all based on the available instantaneous channel state information (CSI) at the base station (BS). 
Therefore, the user in frequency-division duplexing (FDD) systems has to keep on feeding back the CSI to the BS, thereby occupying large uplink transmission resources. 
Recently, deep learning (DL) has achieved great success in the CSI feedback.
However, the existing works just focus on improving the feedback accuracy and ignore the effects on the following modules, e.g., beamforming (BF).
In this paper, we propose a DL-based CSI feedback framework for BF design, called CsiFBnet.
The key idea of the CsiFBnet is to maximize the BF performance gain rather than the feedback accuracy.
We apply it to two representative scenarios: single- and multi-cell systems.
The CsiFBnet-s in the single-cell system is based on the autoencoder architecture, where the encoder at the user compresses the CSI and the decoder at the BS generates the BF vector.
The CsiFBnet-m in the multi-cell system has to feed back two kinds of CSI: the desired and the interfering CSI.
The entire neural networks are trained by an unsupervised learning strategy.
Simulation results show the great performance improvement and complexity reduction of the CsiFBnet compared with the conventional DL-based CSI feedback methods.

\end{abstract}



\section{Introduction}
\label{introduction}

\IEEEPARstart{M}{assive} multiple-input multiple-output (MIMO) is one of the key techniques in future wireless communications \cite{5595728,8808168}.
The massive MIMO systems, equipping base stations (BSs) with a very large number of antennas, have achieved dramatic gains in spectral and energy efficiency as well as to simplify the signal processing \cite{6798744}.
However, these potential benefits can be acquired only when the instantaneous and accurate channel state information (CSI) is available at the transmitter. 
In time-division duplexing (TDD) systems, the BSs can infer the downlink CSI from the uplink CSI utilizing the channel reciprocity.
However, there is weak reciprocity between the downlink and uplink channels in frequency-division duplexing (FDD) systems, which are widely employed by the existing cellular systems.  
Therefore, after estimating the downlink channel from the pilots sent by the BS, the user keeps on feeding CSI back to the BS.
The feedback overhead in codebook-based CSI feedback methods scales linearly with the dimension of CSI matrix, which is determined by the antenna numbers of the BS and user. 
In massive MIMO systems, this strategy is infeasible due to the substantial antennas at the BS, which consumes many precious uplink bandwidth resources\cite{6798744}. 
Therefore, developing a technique that can greatly reduce the feedback overhead while keeping on the feedback accuracy is urgent.

Compressive sensing (CS) has been regarded as a promising technology, which exploits the CSI sparsity in certain domain \cite{8350399}. 
The spatial correlations among nearby antennas are utilized to compress the CSI in spatial-frequency domain for overhead reduction in \cite{6214417}.
The authors in \cite{6816089} consider joint downlink channel estimation and feedback by exploiting the spatially joint sparsity of multiple users' CSI matrices due to the shared local scatters.
The CSI reconstruction at the BS turns into an NP-hard optimization problem and is often solved by iterative algorithms, thereby consuming substantial computing resources and time.
Besides, the sparsity of CSI is the only prior information and the CS-based feedback methods can not exploit the environment information at all.
These shortcomings make the CS-based methods difficult to be implemented in practical systems.
Codebook-based approaches are usually adopted to reduce feedback overhead. However, the feedback quantities resulting from these approaches need to
be scaled linearly with the number of transmit antennas and are prohibitive in a massive MIMO regime \cite{8322184}.
There are some works, e.g., \cite{1237136,4299601}, which quantize and feedback the beamformer with a codebook and use the spectrum efficiency as the optimization objective to select the codeword.

Recently, deep learning (DL) has achieved great success in physical layer communications \cite{8663966,8233654}, e.g., channel estimation \cite{qi2020acquisition},	joint channel estimation and signal detection \cite{8052521}, beamforming (BF) \cite{8935405}, and semantic communication \cite{xie2020deep}.
The authors in \cite{8322184} first apply DL to CSI feedback problem and propose an autoencoder architecture, CsiNet, where the encoder at the user compresses the CSI by fully connectedly (FC), layer and the decoder at the BS reconstructs the CSI from the feedback codeword and then uses convolutional layer to refine the initial reconstructed CSI.
The existing DL-based CSI feedback methods focus on improving the feedback accuracy by introducing expert knowledge and proposing novel NN architectures \cite{00169}.
Long short-term memory architecture is introduced to exploit the temporal correlation in time-varying massive MIMO channels in \cite{8482358}.
The work in \cite{8972904} introduces an offset NN to minimize the nonuniform quantization distortion and proposes a multiple-rate feedback framework.
CoCsiNet and distributed DeepCMC in \cite{guo2020dl,yang2020distributed} exploit the correlation between the nearby users and propose a cooperative and distributed feedback framework.
The computational complexity is considered in \cite{guo2020wcm,cao2020lightweight} and NN quantization and pruning are adopted in \cite{guo2020wcm}.
The authors in \cite{8972904,8625537,9076084} consider the feedback errors and an extra NN, called DNNet, is introduced to improve the performance of channel feedback in \cite{9076084}.
The work in \cite{liu2020adversarial} investigates the effects of adversarial attack on the DL-based CSI feedback.
The practical performance of the DL-based CSI feedback is investigated in \cite{9200894}.
Besides, many novel NN architectures, e.g., CsiNet+ \cite{8972904}, ConvCsiNet \cite{cao2020lightweight}, ConvlstmCsiNet \cite{8951228}, CRNet \cite{lu2019multi}, and ReNet \cite{9126231} are proposed to improve the feedback accuracy.

The goal of the feedback is to acquire the CSI as accurate as possible and the physical meaning of CSI is ignored.
The feedback mechanism of the CS- or DL-based methods is to drop the redundant or unimportant information, which is sometimes based on the sparsity assumption \cite{8972904}.
Reducing the feedback mean-squared error (MSE) has been the direct optimization goal of the most works \cite{8322184,8482358,8972904,guo2020dl,yang2020distributed,guo2020wcm,cao2020lightweight}.
However, the MSE sometimes can not really measure the signal fidelity \cite{4775883}.
In massive MIMO, the most important information in CSI is the phase and magnitude of the paths.
Unfortunately, the MSE function equally treats all information and sometimes even keeps on secondary information at the expense of useful information.
Therefore, although the MSE is lower, the total communication system may perform worse.
This also occurs in the computer vision domain, where how to assess the quality of the reconstructed images is a great challenge.
Signal-to-noise ratio (SNR) and structural similarity (SSIM) are the most widely used assessment metrics.
But, they can not well predict the subjective human perception of image fidelity and quality \cite{4775883,2020perceptual}.
Subjective assessment is the most reliable and accurate.
However, subjective test cannot be directly used as the optimization metric and is expensive and time-consuming.
Fortunately, in the CSI feedback domain, subjective metrics are not needed and we can assess the effects of feedback accuracy on the subsequent communication modules, which require accurate CSI.
Meanwhile, the performance of subsequent communication modules, e.g., BF, can be the optimization goal of the feedback, which can be also regarded as joint feedback and BF design.

Besides, the existing DL-based CSI feedback works only consider 
some simple massive MIMO systems and extending the DL techniques to more complicated scenarios, e.g., multi-cell systems \cite{6879482}, is necessary.
In the single-cell FDD systems, only the downlink channel between the BS and user needs to be fed back.
In the multi-cell systems, the user has to feed back the desired and the interfering channels, respectively, to  combat co-channel interference \cite{5613944}.
Extending DL-based CSI feedback and BF to this complicated scenario is a challenge.

In this paper, we propose a DL-based implicit \underline{CSI} \underline{f}eedback NN framework for \underline{B}F, called CsiFBnet, in massive MIMO systems.
First, we consider a simple scenario, i.e., a single-user and single-cell massive MIMO system.
The proposed framework, called CsiFBnet-s, replaces an end-to-end procedure including CSI compression, quantization, generating BF vectors satisfying the constant modulus constraint.
This framework is based on the autoencoder architecture, in which a uniform quantization is embedded between the encoder and decoder.
Then, we consider a more complicated multi-cell scenario and develop an NN framework, called CsiBFnet-m, to realize CSI feedback for maximizing the sum-rates (at high SNR), adopting a soft hand-off model \cite{4385782}.
Different from the single-cell, the desired and the interference CSI should be fed back to the BS and taken into consideration during BF vector design. 
Our contributions in this paper are summarized as follows:
\begin{itemize}
\item We propose an implicit CSI feedback strategy, CsiFBnet, which aims at improving the performance of BF rather than the CSI feedback accuracy, i.e., the MSE.
\item A single-user and single-cell scenario is taken into consideration.
In the proposed CsiFBnet-s, the encoder at the user side compresses and quantizes the downlink CSI and the decoder at the BS generates the BF vectors satisfying the constant modulus constraint from the feedback measurements.
This method shows great performance improvement especially when the number of feedback bits is extremely constrained.
\item Then, we extend the proposed framework to a more complicated scenario, i.e., the multi-cell massive MIMO system, and adopt the soft hand-off model with a single interferer.
In the multi-cell systems, both of the desired and interference CSI should be fed back to the BS.
In the conventional BS cooperation, the interference CSI is exchanged among the nearby BSs but the backhaul links among BSs are capacity-limited. 
In the proposed CsiFBnet-m, only the compressed and quantized CSI is exchanged among the nearby BSs, thereby reducing the overhead of backhaul. 
\item To maximize the sum-rate of the multi-cell systems, the decoder at the BS generates the BF vectors locally from the compressed feedback measurements, which outperforms the methods of feeding back accurate CSI and designing BF vectors separately.

\end{itemize}

The rest of this paper is organized as follows. 
Section \ref{s2} introduces the single- and multi-cell systems and channel model.
Section \ref{s3} presents the DL-based feedback for BF design in the single-cell system.
Section \ref{s4} extends the proposed feedback framework to the multi-cell system.
Section \ref{s5} provides the numerical results of the proposed methods and evaluate the robustness of the CsiFBnet-m.
Section \ref{s6} finally concludes our paper.

\section{System Model}
\label{s2}
After introducing the single-cell massive MIMO system and channel model, we will describe the multi-cell soft hand-off model.

\subsection{Single-cell Massive MIMO System}
\label{s2-1}
We consider a single-cell massive MIMO system
\footnote{Our proposed method can
be easily extended to more complicated scenario, e.g., wideband scenario and multi-antenna user scenario, by embedding the encoder and decoder for the wideband scenario to the proposed framework.},
where there is a single receiver antenna at the user and $N_{\rm t}$ uniform linear array (ULA) transmit antennas at the BS with one radio frequency (RF) chain
\footnote{As in \cite{8847377}, we assume the system with one RF chain just for simplicity, and the following works can be applied to the system with multiple RF chains by changing the neuron number of the last FC layer.}.
When the BS conveys the symbol $s$ with normalized average symbol energy, i.e., $E\{ |s|^2\}=1$, to the user with a linear BF vector $\mathbf{w} \in \mathbb{C}^{N_{\rm t} \times 1}$, the received signal at the user is expressed as follows:
\begin{equation}
y = \mathbf{h}^{H} \mathbf{x} + n,
\end{equation}
where $\mathbf{h}\in \mathbb{C}^{ N_{\rm t} \times 1 }$ is the downlink channel between the BS and the user, $n$ is the complex additive Gaussian noise (AWGN) with zero mean and variance $\sigma^2$, and $\mathbf{x}\in \mathbb{C}^{N_{\rm t} \times 1}$ is the precoded signal, which is given by $\mathbf{x} = \mathbf{w}s$.
Here, a widely used spatial multi-path channel model is used and $\mathbf{h}$ can be formulated as \cite{8590736}
\begin{equation}
\mathbf{h} = \sum_{c=1}^{N_{\rm c}} \sum_{s=1}^{N_{\rm s}}  g_{c,s} \textbf{a}(\theta_{c,s}),
\end{equation}
where ${N_{\rm c}}$ stands for the number of the scattering clusters, i.e., path number, ${N_{\rm s}}$ in each scattering cluster, $g_{c,s}$ is the complex gain of the $s$-th sub-path in $c$-th  scattering cluster, and $\theta_{c,s}$ is the corresponding azimuth angle-of-departure (AoD) of this sub-path.
When the BS is with ULA transmit antennas, the steering vector $\mathbf{a}(\theta) \in \mathbb{C}^{N_{\rm t} \times 1} $ can be written as
\begin{equation}
\mathbf{a}(\theta)  = [1,  e^{-j 2 \pi \frac{d}{\lambda} sin(\theta)},   \ldots  ,  e^{-j 2 \pi \frac{(N_{\rm t} - 1)d}{\lambda} sin(\theta)}]^T,
\end{equation}
where $d$ and $\lambda$ are the antenna element spacing and carrier wavelength, respectively.

\subsection{Multi-cell Massive MIMO System}
\begin{figure*}[t]
    \centering 
     \includegraphics[scale= 0.85]{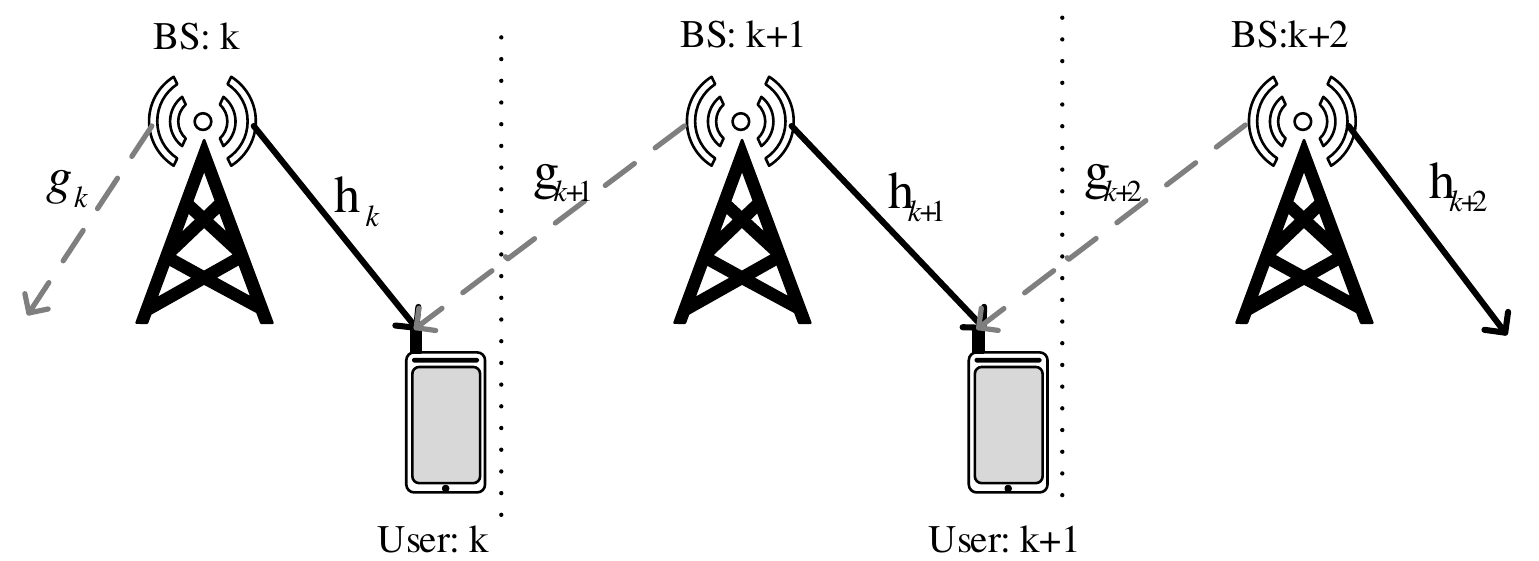}    
	\caption{\label{soft-hand-off}Illustration of the multi-cell soft hand-off model. The solid and the dashed lines represent the desired channel $\mathbf{h}$ and the interfering channel $\mathbf{g}$, respectively.} 
\end{figure*}
The multi-cell massive MIMO system adopted in this paper is the popular soft hand-off model \cite{4385782}, which is a simplified Wyner-type network model \cite{850679} and widely used in multi-cell cooperation problem, e.g., \cite{5613944,jing2008multicell,4729770,6779600,9047177}.
Since our work is the first that considers the DL-based feedback in the multi-cell systems, we use the simple but representative  soft hand-off model.
Different from the single-cell systems, the users in the multi-cell systems have to feed back both the desired and the interfering channels together.
Although the considered multi-cell model is simple, it also includes the most important characteristic of the multi-cell systems for the feedback problem.
Therefore, we think the multi-cell setup is reasonable and can give a guideline for future research.

As shown in Fig. \ref{soft-hand-off}, we consider that the BS with ULA transmit antennas in each cell serves a single active user by intra-cell time division multiple access (TDMA).
This scenario is a common assumption used in multi-cell cooperative works for massive MIMO systems \cite{5613944}.
In this paper, this simple but representative interferer assumption is a good starting to analyze the DL-based CSI feedback for BF design.
The user at the cell center only receives the desired signal from the corresponding cell.
When the user is located at the cell edge, it receives the signals not only from its corresponding cell but also from the adjacent cell.
The second kind of signal is regarded as interference, which seriously affects the user's downlink performance.
Therefore, the BF design for the users at the cell edge should take this interference into consideration rather than only considering the desired signal.

Different from the Wyner model, where the number of cells is assumed to be infinite, we here consider a multi-cell system with $K$ cells.
In the $k^{\rm th}$ cell, the BS only serves the $k^{\rm th}$ user, for $k=1,\ldots,K$.
Similar to the single-cell scenario in Section \ref{s2-1}, each BS in the multi-cell system is equipped with $N_{\rm t}$ ULA antennas and each user supports a single receive antenna, which is a standard multiple-input single-output (MISO) system.
As shown in Fig. \ref{soft-hand-off}, since the $k^{\rm th}$ user receives the signals from both of the $k^{\rm th}$ and $(k+1)^{\rm th}$ BSs, the user should feed back the CSI corresponding to the downlink channel between the user and $k^{\rm th}$ BS and CSI corresponding to the downlink channel between the user and $(k+1)^{\rm th}$ BS, which are denoted by $\mathbf{h}_k\in \mathbb{C}^{ N_{\rm t} \times 1 }$ and $\mathbf{g}_{k+1}\in \mathbb{C}^{ N_{\rm t} \times 1 }$, respectively.
In the soft hand-off multi-cell system, assuming the desired and interfering signal powers received at the $k^{\rm th}$ user side are denoted by $\gamma_{k,(\mathrm{d})}$ and $\gamma_{k,(\mathrm{i})}$, respectively, the received signal $y_{k}$ of $k^{\rm th}$ user is given by
\begin{equation}
y_{k}=\sqrt{\gamma_{k,(\mathrm{d})}} \mathbf{h}_{k}^{T} \mathbf{w}_{k} s_{k}+\sqrt{\gamma_{k,(\mathrm{i})}} \mathbf{g}_{k+1}^{T} \mathbf{w}_{k+1} s_{k+1}+n_{k}
\end{equation}
where $s_{k}$ and $s_{k+1}$ denote the symbol transmitted from the $k^{\rm th}$ and $(k+1)^{\rm th}$ BSs, respectively, $\mathbf{w}_{k}$ and $\mathbf{w}_{k+1}$ are the BF vectors of the $k^{\rm th}$ and $(k+1)^{\rm th}$ BSs, respectively, and $n_k$ is the zero-mean AWGN noise of the $k^{\rm th}$ user with $ \mathbb{E}\{ |n_k|^2\}=N_{\rm 0}$.

\section{DL-based CSI feedback for BF in single-cell systems}
\label{s3}
In this section, we consider the CSI feedback for BF in the single-cell massive MIMO system.
Unlike the conventional DL-based digital CSI feedback methods, which only focus on decreasing the feedback $\rm NMSE$, the CSI feedback in this part considers the effects on the subsequent module, i.e., BF module.
The key idea is to replace the MSE loss function with the spectral efficiency.
First, we will briefly introduce the BF design in the single-cell system.
Then, the DL-based CSI feedback framework for BF is proposed.
Details about the NNs in the CsiFBnet-s are given at last.

\subsection{BF Design for Single-cell System}

In the BF design problem, spectral efficiency is widely used to assess the BF performance as in \cite{5613944,8847377,8935405}.
The spectral efficiency of the first studied system, i.e., single-cell and single-user system, can be formulated as 
\begin{equation}
\label{se4s}
R = {\rm{log_2}}\Big(1+\frac{\left|\mathbf{h}^{H} \mathbf{w}  \right|^2}{\sigma ^2}\Big).
\end{equation}
Since the symbol is multiplied by a scalar
digital precoder $v_{\rm D}$ and an analog precoder $\mathbf{v}_{\rm RF}\in \mathbb{C}^{N_{\rm t} \times 1}$ using phase shifters, respectively, the BF vector $\mathbf{w}$ can written as 
\begin{equation}
\mathbf{w} = \mathbf{v}_{\rm RF} v_{\rm D}.  
\end{equation}
If there is a constant modulus constraint on
analog BF, i.e., $\left|[\mathbf{v}_{\rm RF}]_{i}\right|^2 = 1$ and a maximum transmit power constraint, i.e., $|\mathbf{w}|^2 \le P$, the optimal $v_{\rm D}$ for maximizing $R$ is given by $\sqrt{P/{N_{\rm t}}}$.
Therefore, the BF vector $\mathbf{w}$ optimization problem turns into optimizing $\mathbf{v}_{\rm RF}$, which is given by
\begin{equation}
\label{se4single}
\begin{array}{c}
\underset{\mathbf{v}_{\mathrm{RF}}}{\operatorname{maximize}} \log _{2}\left(1+\frac{\rho}{N_{\mathrm{t}}}\left\|\mathbf{h}^{H} \mathbf{v}_{\mathrm{RF}}\right\|^{2}\right) \\
\text { subject to }\left| \left[\mathbf{v}_{\mathrm{RF}}\right]_{i}\right|^{2}=1, 
\text { for } i=1, \ldots,N_{\mathrm{t}},
\end{array}
\end{equation}
where $\rho = \frac{P}{\sigma ^2}$ represents the signal-to-noise ratio (SNR), which is independent of the BF vectors and periodically fed back to the BS in the form of channel quality
indicators (CQI) in 5G New Radio (NR) \cite{8666153} for scheduling and adaptive modulation and coding purposes, so as to compensate for the channel impairments and also to reduce multiple re-transmissions.
To save the feedback bandwidth, CQI should be quantized and then fed back to the BS. It has been shown in \cite{4510711} that SNR quantization does not affect the sum-rates of a single-cell multiuser MIMO system significantly. Hence, we assume that the BS has known the perfect CQI in this work.

\subsection{Prior Art }

The existing DL-based CSI feedback methods are based on the autoencoder architecture, which is widely used in DL-based image compression area.
The encoder at the user reduces the dimension of the downlink CSI by reducing the neuron number of FC layers, e.g., \cite{8322184,8482358,8972904,guo2020dl}, or using the pooling layer, e.g.,\cite{cao2020lightweight,yang2020distributed}.
The decoder at the BS reconstructs the downlink CSI from the feedback measurement of bits by increasing the neuron number of FC layers or using stacked deconvolution or upsampling layers.
Since the user should feed back the bits rather than analog numbers to the BS, the encoder is usually followed by a quantization module.
The entire feedback process can be formulated as
\begin{equation}
\hat{\mathbf{h}} = {\rm f_{de}} \Big(  \mathcal{Q} \big(  {\rm f_{en}}(\mathbf{h},\Theta_{\rm en}) \big),\Theta_{\rm de} \Big),
\end{equation}
where $ {\rm f_{en}(\cdot)}$ and $ {\rm f_{de}(\cdot)}$ denote the compression and the reconstruction operations, respectively, $\Theta_{\rm en}$ and  $\Theta_{\rm de}$ are the NN weights at the encoder and the decoder, respectively, and $ \mathcal{Q}(\cdot)$ represents the quantization operation following the compression module.

The NN training goal is to minimize the MSE between the original CSI $\mathbf{h}$ at the user and the obtained CSI $\hat{\mathbf{h}}$ at the BS by updating the NN weights as
\begin{equation}
(\hat \Theta_1, \hat \Theta_2) 
= \mathop{\arg\min}_{\Theta_1, \Theta_2} \ \ \Big \|  \mathbf{h}-   
{\rm f_{de}} \big(  \mathcal{Q} \big(  {\rm f_{en}}(\mathbf{h},\Theta_{\rm en}) \big),\Theta_{\rm de} \Big) \Big\|_2^2,
\end{equation}
where $\| \cdot  \|_2^2$ represents the Euclidean norm.
The final feedback overhead, i.e., the number feedback bits, is decides by the compression ratio and quantization bits as
\begin{equation}
\label{crEq}
N_{\rm bits} = \frac{L\times B}{\beta},
\end{equation}
where $L$ is the original dimension of CSI, $ \beta >1$ denotes the compression ratio, and $B$ is the quantization bit.
The most widely used metrics in the DL-based CSI feedback is
normalized MSE (NMSE) as
\begin{equation}
{\rm NMSE} = {\rm E} \bigg \{  \frac{  { \Vert  \hat{\textbf{h}} - \textbf{h} \Vert }_2^2  }{ { \Vert   \textbf{h} \Vert }_2^2 }   \bigg \}.
\end{equation}

\subsection{Framework of The Proposed CsiFBnet-s}

\begin{figure*}[t]
    \centering 
     \includegraphics[scale= 0.75]{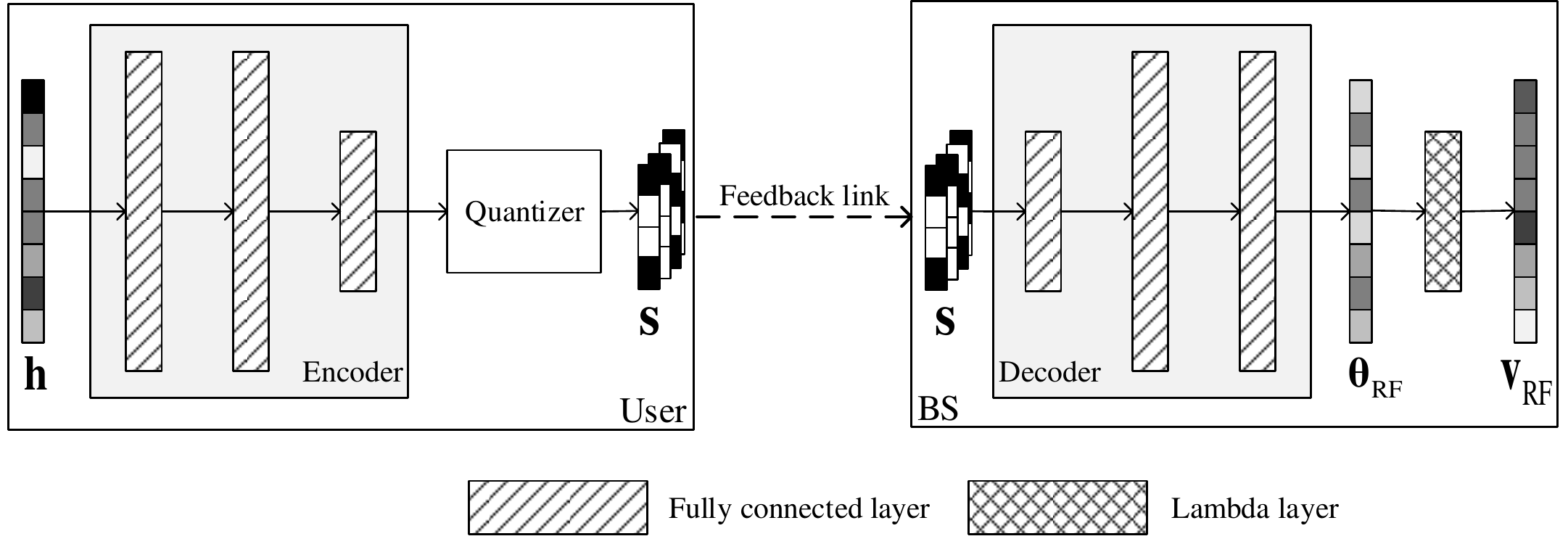}    
	\caption{\label{CsiFBnet_s}Illustration of the proposed CsiFBnet-s for the single-cell massive MIMO system. The input is the downlink CSI and the output is the analog BF vector $\mathbf{v}_{\rm RF}$.} 
\end{figure*}

In the FDD massive MIMO system, the user first compresses the estimated CSI and then quantizes the measurement vector to reduce the feedback overhead.
Once the BS obtains the feedback bits, it designs the analog BF vector, $\mathbf{v}_{\rm RF}$, using the information from the feedback, i.e., the quantized measurement vector.
The main challenge for the CSI feedback is how to choose the information, which is necessary for the BF design. 
The end-to-end optimization characteristic of the DL technique provides a potential way to tackle this challenge.

As shown in Fig. \ref{CsiFBnet_s}, the proposed CsiFBnet-s framework contains three main modules, i.e., encoder, quantizer, and decoder.
The encoder and the decoder are both consisted of FC layers.
The first two FC layers at the encoder extract the CSI features and the last FC layer with $L/\gamma$ neurons compresses the downlink CSI by $\gamma$ times.
The compressed CSI is quantized using a uniform quantizer to generate feedback codeword $\mathbf{s}$. 
Different from the conventional feedback strategy, which should first reconstruct the CSI at the BS, the decoder in the proposed CsiFBnet-s directly generates the analog vector $\mathbf{v}_{\rm RF}$ from the feedback codeword, $\mathbf{s}$.

As shown in (\ref{se4single}), there is a constant modulus constraint on the analog BF vector $\mathbf{v}_{\rm RF}$, that is, $\left| \left[\mathbf{v}_{\mathrm{RF}}\right]_{i}\right|^{2}=1, \text { for } i=1, \ldots,N_{\mathrm{t}}$. 
However, the output of the FC layer with or without an activation operation can not satisfy the constraint on  the complex-valued $\mathbf{v}_{\rm RF}$.
Inspired by \cite{8847377}, we add an extra Lambda layer following the last FC layer at the decoder.
Let $\boldsymbol{\theta}_{\rm RF}$ represents the real-valued output of the last FC layer, whose elements are all in the range of $(- \infty,+ \infty)$ and correspond to the phases of the analog BF coefficients in the physical meaning, the output of the Lambda layer is defined as
\begin{equation}
\mathbf{v}_{\rm RF} = {\rm exp}({{\rm j}\cdot\boldsymbol{\theta}_{\rm RF} }) =  {\rm cos}(\boldsymbol{\theta}_{\rm RF}) +  {\rm j}\cdot  {\rm sin}(\boldsymbol{\theta}_{\rm RF}),
\end{equation}
where $\rm j = \sqrt{-1}$.
Thus, the entire process of the proposed CsiFBnet-s framework
can be formulated as
\begin{equation}
\mathbf{v}_{\rm RF} = {\rm exp}\bigg( {{\rm j}\cdot{{\rm f_{de,s}} \Big(  \mathcal{Q} \big(  {\rm f_{en,s}}(\mathbf{h},\Theta_{\rm en,s}) \big),\Theta_{\rm de,s} \Big)}}\bigg),
\end{equation}
where $ {\rm f_{en,s}(\cdot)}$ and $ {\rm f_{de,s}(\cdot)}$ denote the compression and the BF generation operations in the  CsiFBnet-s framework, respectively, and $\Theta_{\rm en,s}$ and  $\Theta_{\rm de,s}$ are the NN weights at the encoder and the decoder, respectively.
To maximize the spectral efficiency in (\ref{se4s}), the loss function is defined as
\begin{equation}
{\rm Loss_s} = -    \mathbf{h}^{H} {\rm exp}\bigg( {{\rm j}\cdot{{\rm f_{de,s}} \Big(  \mathcal{Q} \big(  {\rm f_{en,s}}(\mathbf{h},\Theta_{\rm en,s}) \big),\Theta_{\rm de,s} \Big)}}\bigg).
\end{equation}
Different from the conventional DL-based CSI feedback, which has the ground truth, i.e., label, the CsiFBnet-s has no label, that is, predefined optimal $\mathbf{v}_{\rm RF}$.
Therefore, the training of the CsiFBnet-s belongs to data-driven unsupervised leaning \cite{8935405}.

\subsection{NN Details of the CsiFBnet-s}
\begin{table*}[ht]
\caption{\label{NN1}NN details in the proposed CsiFBnet-s.}
\centering
\resizebox{\textwidth}{!}{
\begin{threeparttable}
\begin{tabular}{c|ccccc}
\hline \hline
                         & Layer name   & Output shape        & Activation operation       & Parameter number                        & FLOPs                                    \\ \hline \hline
\multirow{4}{*}{Encoder} & Input        & $2N_{\rm t}\times1^*$ & \textbackslash{} & 0                                       & 0                                        \\
                         & FC1          & $2N_{\rm t}\times1$ & Leaky ReLu       & $2N_{\rm t}\times(2N_{\rm t}+1)$        & $(4N_{\rm t}-1)\times 2N_{\rm t}$        \\
                         & FC2          & $2N_{\rm t}\times1$ & Leaky ReLu       & $2N_{\rm t}\times(2N_{\rm t}+1)$        & $(4N_{\rm t}-1)\times 2N_{\rm t}$        \\
                         & FC3          & $2N_{\rm t}/\beta \times 1 $   & Tanh             & $(2N_{\rm t}/\beta)\times(2N_{\rm t}+1)$ & $(4N_{\rm t}-1)\times (2N_{\rm t}/\beta)$ \\  \hline \hline
\multirow{4}{*}{Decoder} & FC4          & $4N_{\rm t}\times1$ & Leaky ReLu       & $4N_{\rm t}\times(2N_{\rm t}/\beta+1)$   & $(4N_{\rm t}/\beta-1)\times 4N_{\rm t}$   \\
                         & FC5          & $2N_{\rm t}\times1$ & Leaky ReLu       & $2N_{\rm t}\times(4N_{\rm t}+1)$        & $(8N_{\rm t}-1)\times 2N_{\rm t}$        \\
                         & FC6          & $N_{\rm t}\times1$  & \textbackslash{} & $N_{\rm t}\times(2N_{\rm t}+1)$         & $(4N_{\rm t}-1)\times N_{\rm t}$         \\
                         & Lambda layer & $N_{\rm t}\times1$  & \textbackslash{} & 0                                       & 0                                       \\\hline \hline
\end{tabular}
\begin{tablenotes}
        \footnotesize
        \item[*] The real and the imaginary parts of the CSI are concatenated and thus, the input vector shape is $2N_{\rm t}\times 1$.
        \item[**] The quantizer module, following the encoder, is neglected. 
        \item[***] The FLOPs of the activation operation and the lambda layer is neglected since they are much smaller than those of FC layers.
 \end{tablenotes}
\end{threeparttable}
}
\end{table*}

In this subsection, we introduce the details about CsiFBnet-s including the NN architecture, weight number, and floating point operations (FLOPs).
Since this paper is not focusing on proposing novel NN architectures to improving feedback performance, the NNs in the CsiFBnet-s are based on the vanilla FC layers.
As shown in Table \ref{NN1}, the neuron number of the FC layer is not fixed but depends on the CSI dimension.
If the CSI dimension increases, the NNs will naturally be more complicated.
The activation function of the FC3 layer is ${\rm Tanh}$, whose outputs are normalized into the range of $(-1,+1)$. 
Then, the uniform quantizer discretize the output of the FC3 layer by 4 bits, following the setting in \cite{guo2020dl,8845636}.
Since the $\rm round(\cdot)$ operation in the quantization module is non-differentiable, the gradient in the back-forward process can not pass through this module.
Therefore, we set its gradient to a unit constant, thereby making it possible to train the entire NN framework, including the encoder, the quantization module, and the decoder, in an end-to-end way. 

To show the NN complexity, we calculate the NN weight number and FLOPS in this subsection.
According \cite{molchanov2016pruning}, the weight number and FLOPs of FC layers are calculated by
\begin{equation}
N_{\rm FC} = O \times (I+1),
\end{equation}
\begin{equation}
{\rm FLOPs_{FC}} = O \times (2I-1),
\end{equation}
where $I$ and $O$ are the input and the output dimensions, respectively.
Therefore, the total NN weight number of the proposed CsiFBnet-s is $(18 + 12/\beta){N_{\rm t}^2} + (11 + 2/ \beta){N_{\rm t}}$ and the total FLOPs number is $(36 + 24/\beta){N_{\rm t}^2} - (11 + 2/ \beta){N_{\rm t}}$.

\begin{figure}[t]
    \centering 
     \includegraphics[scale= 0.75]{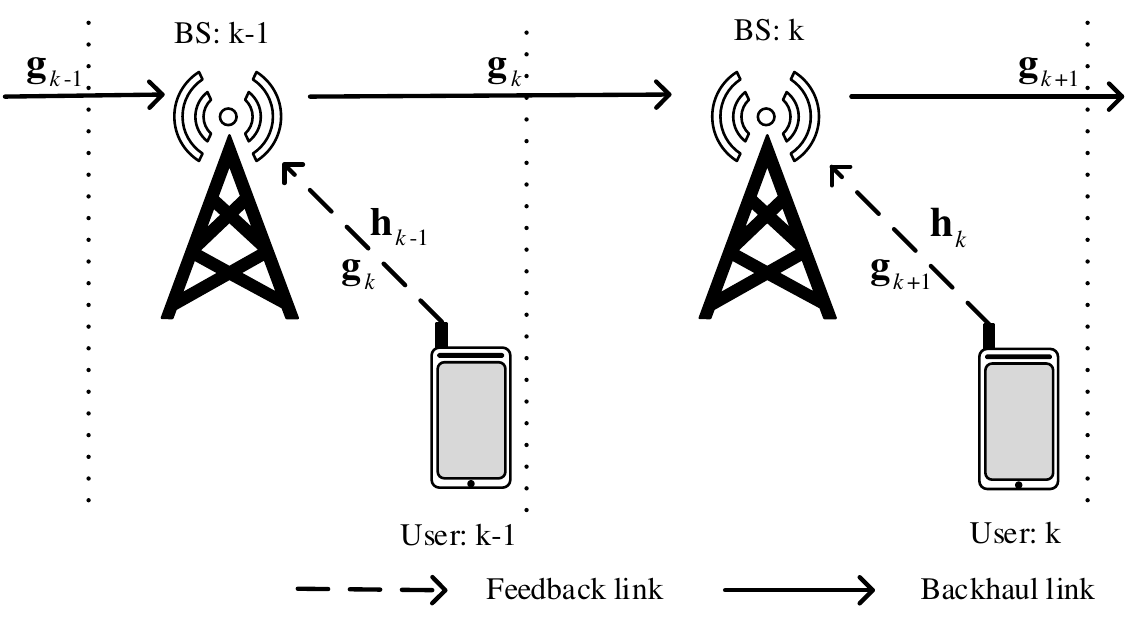}    
	\caption{\label{multi-cell-exchange}Illustration of CSI feedback and exchange in the multi-cell soft hand-off system. 
The $k^{\rm th}$ user feeds back its obtained CSI including the desired and the interfering channel, i.e., $\mathbf{h}_k$ and $\mathbf{g}_{k+1}$, to the correspond $k^{\rm th}$ BS through the uplink, which leads to a large overhead.
The $(k-1)^{\rm th}$ BS sends $\mathbf{g}_{k}$ to the $k^{\rm th}$ BS through a backhaul link.} 
\end{figure}

\section{DL-based CSI feedback for BF in multi-cell systems}
\label{s4}
In this section, we consider the CSI feedback for BF in the multi-cell soft hand-off system.
First, we introduce the BF design for the soft hand-off system.
Then, the CsiFBnet-m framework are present.
The details of the NNs in the CsiFBnet-m are given at last.

\begin{figure*}[t]
    \centering 
     \includegraphics[scale= 0.55]{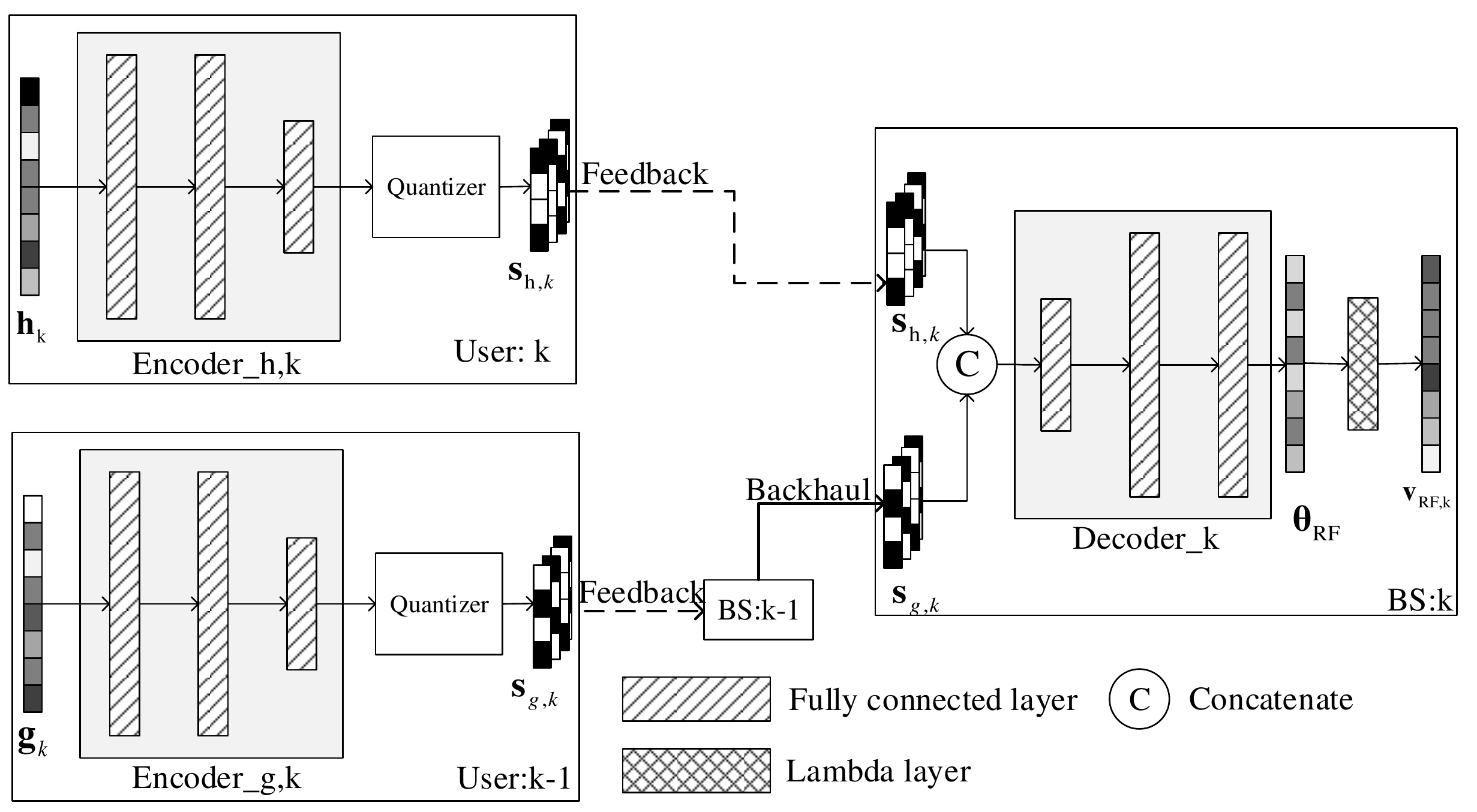}    
	\caption{\label{CsiFBnet_m}Illustration of the Proposed CsiFBnet-m for the soft hand-off multi-cell massive MIMO system. There are two encoders and one decoder at $k^{\rm th}$ and $(k-1)^{\rm th}$ users, and $k^{\rm th}$ BS, respectively. } 
\end{figure*}

\subsection{BF Design for Multi-cell System}

The signal-to-interference-noise ratio (SINR) of the $k^{\rm th}$ user in the soft hand-off system can be expressed as 
\begin{equation}
\label{snrEq}
\begin{aligned}
\mathrm{SINR}_{k} &= \frac{\gamma_{k,(\mathrm{d})} |\mathbf{h}_k^H \mathbf{w}_k|^2 }{\gamma_{k,(\mathrm{i})} \left|\mathbf{g}_{k+1}^H \mathbf{w}_{k+1}\right||^2+N_{\rm 0}}\\
&=\frac{\left|\mathbf{h}_{k}^{H} \mathbf{w}_{k}\right|^{2}}{\alpha_{k}\left|\mathbf{g}_{k+1}^{H} \mathbf{w}_{k+1}\right|^{2}+\frac{1}{\rho_{k},(\mathrm{d})}},
\end{aligned}
\end{equation}
where $\alpha_{k} = \gamma_{k,(\mathrm{i})} / \gamma_{k,(\mathrm{d})} \in [0,1] $ represents the strength ratio between the interfering and the desired signal, and $\rho_{k},(\mathrm{d}) = N_{\rm 0} / \gamma_{k,(\mathrm{d})}$ denotes the SNR of the $k^{\rm th}$ user, which is fed back in the form of CQI.
The parameter $\alpha_k$ does not need to be fed back.
The $k$-th BS can obtain the desired and interfering channels by the feedback and backhaul links, respectively.
The sum-rate of all users in the multi-cell system is given by
\begin{equation}
\label{sum-rate}
\begin{aligned}
R_{\rm sum} &= \sum_{k} {\rm{log_2}}(1+{\rm SINR}_{k})\\
&= \sum_{k} {\rm{log_2}}\Big(1+\frac{\left|\mathbf{h}_{k}^{H} \mathbf{w}_{k}\right|^{2}}{\alpha_{k}\left|\mathbf{g}_{k+1}^{H} \mathbf{w}_{k+1}\right|^{2}+\frac{1}{\rho_{k},(\mathrm{d})}}\Big).
\end{aligned}
\end{equation}

From (\ref{sum-rate}), the design of BF vector $\mathbf{w}_{k}$ greatly influences the $\mathrm{SINR}_{k}$, thereby influencing the sum-rate of the multi-cell system.
Therefore, the BF vector $\mathbf{w}_k$ design turns in to an optimization problem as
\begin{equation}
\label{multicellOptimization}
\underset{\{\mathbf{w}_{k}\}_{k=1}^{K}}{\operatorname{maximize}} \sum_{k} {\rm{log_2}}\Big(1+\frac{\left|\mathbf{h}_{k}^{H} \mathbf{w}_{k}\right|^{2}}{\alpha_{k}\left|\mathbf{g}_{k+1}^{H} \mathbf{w}_{k+1}\right|^{2}+\frac{1}{\rho_{k},(\mathrm{d})}}\Big)
\end{equation}
The $\mathrm{SINR}_{k}$ of the $k^{\rm th}$ user depends on the BF vectors of the $k^{\rm th}$ and the $(k+1)^{\rm th}$ cells, thereby making maximizing the sum-rate turn into a complicated joint optimization across all users in the multi-cell system.
In this case, the BSs should know all channel information, which is impossible and difficult to optimize.
In this paper, following the approximation in \cite{5613944}, a high SINR approximation is adopted to remove the above interdependency among all users, thereby making it a local maximization problem.
When the SINR is high, we can assume ${\rm log_2}(1+{\rm SINR}_k) \approx {\rm log_2}({\rm SINR}_k)$ and thereby the sum-rate, $R_{\rm sum} $, can be written as
\begin{equation}
\begin{aligned}
R_{\rm sum} &\approx \sum_{k} {\rm{log_2}}(\mathrm{SINR}_{k})\\
&={\rm log_2}\Big( \prod_{k} {\rm SINR}_{k}\Big)\\
&={\rm log_2}\Big( \prod_{k} {\frac{\left|\mathbf{h}_{k}^{H} \mathbf{w}_{k}\right|^{2}}{\alpha_{k}\left|\mathbf{g}_{k+1}^{H} \mathbf{w}_{k+1}\right|^{2}+\frac{1}{\rho_{k},(\mathrm{d})}}}\Big).\\
\end{aligned}
\end{equation}
Therefore, the problem of maximizing sum-rate in (\ref{sum-rate}) at high SINRs turns into maximizing the product of SINRs at all users as
\begin{equation}
\label{eqHigh}
\begin{array}{c}
\underset{\{\mathbf{w}_{k}\}_{k=1}^{K}}{\operatorname{maximize}} \quad {\rm log_2}\Big( \prod_{k} {\rm SINR}_{k}\Big) = 
{\rm log_2}\Big( \prod_{k} {\frac{\left|\mathbf{h}_{k}^{H} \mathbf{w}_{k}\right|^{2}}{\alpha_{k}\left|\mathbf{g}_{k+1}^{H} \mathbf{w}_{k+1}\right|^{2}+\frac{1}{\rho_{k},(\mathrm{d})}}}\Big)\\
\begin{split}
\text { subject to }
\mathbf{w}_k = \mathbf{v}_{{\rm RF},k} v_{{\rm D},k},
\left| \left[\mathbf{v}_{{\rm{RF}},k}\right]_{i}\right|^{2}=1, \\
\text { for } k=1,\ldots,K,\quad i=1, \ldots,N_{\mathrm{t}},
\end{split}
\end{array}
\end{equation}
where $ v_{{\rm D},k}$ and $\mathbf{v}_{{\rm RF},k}$ are the scalar digital precoder and analog precoder of the $k^{\rm th}$ cell, respectively.
As in the single-cell system, there is a constant modulus constraint on the analog BF vector, i.e., $\left| \left[\mathbf{v}_{{\rm{RF}},k}\right]_{i}\right|^{2}=1$.
Exploiting the commutativity of the multiplication operation, 
(\ref{eqHigh}) can be expressed as 
\begin{equation}
\begin{array}{c}
\underset{\{\mathbf{w}_{k}\}_{k=1}^{K}}{\operatorname{maximize}} \quad  
{\rm log_2}\Big( \prod_{k} {\frac{\left|\mathbf{h}_{k}^{H} \mathbf{w}_{k}\right|^{2}}{\alpha_{k-1}\left|\mathbf{g}_{k}^{H} \mathbf{w}_{k}\right|^{2}+\frac{1}{\rho_{k-1},(\mathrm{d})}}}\Big)\\
\begin{split}
\text { subject to }
\mathbf{w}_k = \mathbf{v}_{{\rm RF},k} v_{{\rm D},k},
\left| \left[\mathbf{v}_{{\rm{RF}},k}\right]_{i}\right|^{2}=1, \\
\text { for } k=1,\ldots,K,\quad i=1, \ldots,N_{\mathrm{t}}.
\end{split}
\end{array}
\end{equation}
So far, the complicated joint maximization problem in (\ref{multicellOptimization}) turns into a local optimization problem through the high SINR approximation and commutativity of the multiplication operation.
Then, the $k^{\rm th}$ BS designs the BF vector $\mathbf{w}_k$ by
\begin{equation}
\label{localProblem}
\begin{array}{c}
\underset{ \mathbf{w}_{k} }{\operatorname{maximize}} \quad  
{\rm log_2}\Big(   {\frac{\left|\mathbf{h}_{k}^{H} \mathbf{w}_{k}\right|^{2}}{\alpha_{k-1}\left|\mathbf{g}_{k}^{H} \mathbf{w}_{k}\right|^{2}+\frac{1}{\rho_{k-1},(\mathrm{d})}}}\Big)\\
\text { subject to }
\mathbf{w}_k = \mathbf{v}_{{\rm RF},k} v_{{\rm D},k},
\left| \left[\mathbf{v}_{{\rm{RF}},k}\right]_{i}\right|^{2}=1, 
\text { for } i=1, \ldots,N_{\mathrm{t}}.
\end{array}
\end{equation}

\subsection{Framework of The Proposed CsiFBnet-m}
\label{ss3}

Different from the single-cell single-user scenario in Section \ref{s3}, the feedback and the BF design in the soft hand-off multi-cell system need the information of the desired and the interfering channels.
Therefore, as shown in Fig.\ref{multi-cell-exchange}, the $k^{\rm th}$ user feeds back the desired channel $\mathbf{h}_k$ and interfering channel $\mathbf{g}_{k+1}$ to its correspond cell, i.e., the $k^{\rm th}$ BS.
The $k^{\rm th}$ BS obtains the channel $\mathbf{g}_{k}$ from the $(k-1)^{\rm th}$ BS through a backhaul link, which is used for the information exchange between BSs and is assumed to be error-free and without time delay.
Then, the $k^{\rm th}$ BS designs the BF vector by maximizing the function in (\ref{localProblem}) with $\mathbf{h}_k$ and $\mathbf{g}_{k}$.

Correspondingly, as shown in Fig. \ref{CsiFBnet_m}, in the proposed CsiFBnet-m system, there are two NN-based encoders at each user, that is, ${\rm f}_{\rm en,h}^{(k)}$ and ${\rm f}_{\rm en,g}^{(k)}$, which compress the desired and the interfering channels, respectively.
The elements in the compressed channels are then quantized by 4 bits using the uniform quantizer to generate the feedback codewords $\mathbf{s}_{{\rm h},k}$ and $\mathbf{s}_{{\rm g},k}$, respectively.
Once the $k^{\rm th}$ BS obtains the codewords from the feedback and the backhaul links, the decoder ${\rm f}_{\rm de,m}^{(k)}$ generates the analog BF vector $\mathbf{v}_{{\rm{RF}},k}$ to maximize the system sum-rate.
The entire framework can be formulated as
\begin{equation}
\hspace{-0.1cm}\mathbf{v}_{{\rm RF},k} = {\rm exp}\bigg( {{\rm j}\cdot{{\rm f}_{\rm de,m}^{(k)}
 \Big(  \mathcal{Q} \big(  {\rm f}_{\rm en,h}^{(k)}(\mathbf{h}_k,\Theta_{\rm en,h}^{(k)}) \big),
 \mathcal{Q} \big(  {\rm f}_{\rm en,g}^{(k)}(\mathbf{g}_k,\Theta_{\rm en,g}^{(k)}) \big),
\Theta_{\rm de,m}^{(k)} \Big)}}\bigg),
\end{equation}
where $\Theta_{\rm en,h}^{(k)}$ and $\Theta_{\rm en,g}^{(k)}$ are the NN weights at the encoders ${\rm f}_{\rm en,h}^{(k)}$ and ${\rm f}_{\rm en,g}^{(k)}$, and $\Theta_{\rm de,m}^{(k)}$ is the NN weights at the decoder ${\rm f}_{\rm de,m}^{(k)}$. 
Similar to the proposed CsiFBnet-s, there are no ground truth BF vectors and the NNs in the CsiFBnet-m are end-to-end trained by unsupervised learning.
To maximize the sum-rate, the loss function is defined as
\begin{equation}
\label{loss2}
\begin{array}{c}
{\rm Loss_m} = -   {\frac{\left|\mathbf{h}_{k}^{H} \mathbf{w}_{k}\right|^{2}}{\alpha_{k-1}\left|\mathbf{g}_{k}^{H} \mathbf{w}_{k}\right|^{2}+\frac{1}{\rho_{k-1},(\mathrm{d})}}}\\
\text { subject to }
\mathbf{w}_k = \mathbf{v}_{{\rm RF},k} v_{{\rm D},k},
\left| \left[\mathbf{v}_{{\rm{RF}},k}\right]_{i}\right|^{2}=1, 
\text { for } i=1, \ldots,N_{\mathrm{t}}.
\end{array}
\end{equation}

\subsection{NN Details of the CsiFBnet-m}

\begin{table*}[t]
\caption{\label{NN2}NN details in the proposed CsiFBnet-m.}
\centering
\resizebox{\textwidth}{!}{
\begin{threeparttable}
\begin{tabular}{c|ccccc}
\hline \hline
                            & Layer name   & Output shape                                      & Activation       & Parameter number                                                      & FLOPs                                                                                  \\ \hline \hline
\multirow{4}{*}{Encoder\_h} & Input        & $2N_{\rm t}\times1$                               & \textbackslash{} & 0                                                                     & 0                                                                                      \\
                            & FC1h         & $2N_{\rm t}\times1$                               & Leaky ReLu       & $2N_{\rm t}\times(2N_{\rm t}+1)$                                      & $(4N_{\rm t}-1)\times 2N_{\rm t}$                                                      \\
                            & FC2h         & $2N_{\rm t}\times1$                               & Leaky ReLu       & $2N_{\rm t}\times(2N_{\rm t}+1)$                                      & $(4N_{\rm t}-1)\times 2N_{\rm t}$                                                      \\
                            & FC3h         & $2N_{\rm t} / \beta_{\rm h} \times 1$            & Tanh             & $(2N_{\rm t} / \beta_{\rm h} )\times(2N_{\rm t}+1)$                  & $(4N_{\rm t}-1)\times (2N_{t}/\beta_{\rm h})$                                         \\ \hline \hline
\multirow{4}{*}{Encoder\_g} & Input        & $2N_{\rm t}\times1$                               & \textbackslash{} & 0                                                                     & 0                                                                                      \\
                            & FC1g         & $2N_{\rm t}\times1$                               & Leaky ReLu       & $2N_{\rm t}\times(2N_{\rm t}+1)$                                      & $(4N_{\rm t}-1)\times 2N_{\rm t}$                                                      \\
                            & FC2g         & $2N_{\rm t}\times1$                               & Leaky ReLu       & $2N_{\rm t}\times(2N_{\rm t}+1)$                                      & $(4N_{\rm t}-1)\times 2N_{\rm t}$                                                      \\
                            & FC3g         & $2N_{\rm t} / \beta_{\rm g} \times 1$            & Tanh             & $(2N_{\rm t} / \beta_{\rm g} )\times(2N_{\rm t}+1)$                  & $(4N_{\rm t}-1)\times (2N_{t}/\beta_{\rm g})$                                         \\  \hline \hline
\multirow{5}{*}{Decoder\_m} & Concat layer & $2N_{\rm t}(1/\beta_{\rm h} + 1/\beta_{\rm g})$ & \textbackslash{} & 0                                                                     & 0                                                                                      \\
                            & FC4          & $4N_{\rm t}\times1$                               & Leaky ReLu       & $4N_{\rm t}\times(2N_{\rm t}(1/\beta_{\rm h} + 1/\beta_{\rm g})+1)$ & $ (4N_{\rm t}(1/\beta_{\rm h} + 1/\beta_{\rm g})-1)\times 4N_{\rm t}$ \\
                            & FC5          & $2N_{\rm t}\times1$                               & Leaky ReLu       & $2N_{\rm t}\times(4N_{\rm t}+1)$                                      & $(8N_{\rm t}-1)\times 2N_{\rm t}$                                                      \\
                            & FC6          & $N_{\rm t}\times1$                                & \textbackslash{} & $N_{\rm t}\times(2N_{\rm t}+1)$                                       & $(4N_{\rm t}-1)\times N_{\rm t}$                                                       \\
                            & Lambda layer & $N_{\rm t}\times1$                                & \textbackslash{} & 0                                                                     & 0                                                                                     
\\\hline \hline
\end{tabular}
\end{threeparttable}
}
\end{table*}

In this subsection, we introduce the details about CsiFBnet-m, including the NN architecture, weight number, and FLOPs.
The user in the soft hand-off multi-cell system should feed back two kind CSI information to the correspond BS, which causes a feedback-bit allocation problem.
Specifically, the length sum of the codewords $\mathbf{s}_{{\rm h},k}$ and $\mathbf{s}_{{\rm g},k}$ is the final feedback overhead.
How to allocate the feedback bits on the feedback of the desired channel $\mathbf{h}$ and $\mathbf{g}$ is a great challenge \cite{5613944}.
Since the feedback-bit allocation problem is out of the scope of this paper, we define $\beta_{\rm h}$ and $\beta_{\rm g}$ to be the compression ratios of two channels, $\mathbf{h}$ and $\mathbf{g}$ and we will show the effects of the feedback-bit allocation on the system sum-rate in  the following section.

As shown in Table \ref{NN2}, the NN architecture of the CsiFBnet-m is similar to that of the CsiFBnet-s.
Compared with CsiFBnet-s, there is an extra concat layer, which concatenates the output of two encoders, i.e.,  $\mathbf{s}_{{\rm h},k}$ and $\mathbf{s}_{{\rm g},k}$.
Naturally, the  complexity of the CsiFBnet-m is higher than that of CsiFBnet-s, which is caused by the extra encoder and the more complicated FC4 layer.
The total NN weight number of the proposed CsiFBnet-m is $(26 + 12/\beta_{\rm h} + 12/\beta_{\rm g}){N_{\rm t}^2} + (15 + 2/\beta_{\rm h} + 2/\beta_{\rm g}){N_{\rm t}}$ and the total FLOPs number is $(52 + 24/\beta_{\rm h} + 24/\beta_{\rm g}){N_{\rm t}^2} - (15 + 2/\beta_{\rm h} + 2/\beta_{\rm g}){N_{\rm t}}$.
Since the NNs are totally based on the FC layers, the parameter numbers are very huge.
The high requirements of the storage and computational power have been regarded as a main challenge for the DL-based communication algorithms.
As mentioned in \cite{guo2020wcm}, we can apply NN compression and acceleration techniques, e.g., NN quantization and pruning, to the CsiFBnet.

\section{Simulation Results and Discussions}
\label{s5}
To evaluate the performance of our proposed CsiFBnet framework in the single- and multi-cell systems, numerical simulations are conducted.
In this section, we first present the simulation setting, including CSI generation method and the NN training details, e.g., the hyper-parameters. 
Then, we give the spectral efficiency performance of the proposed CsiFBnet-s framework and compare it with the way that does not consider BF design when feedbacking. 
Finally, we introduce the sum-rate performance of the multi-cell system under the CsiFBnet framework, give a comparison with the baselines, and show the effects of the feedback-bit allocation on the desired and the interfering channel feedback.

\subsection{Simulation Setting}
\subsubsection{Channel generation setting}
The 3GPP spatial channel model (SCM) \cite{doi:10.1002/ett.928} is adopted to generate the downlink channel of the urban microcell.
Following the setting in \cite{8590736}, the downlink frequency is 2.17 GHz and the inter-antenna space is set as half wavelength, i.e., $d = c/(2f_{\rm 0})$, where $c$ and $f_{\rm 0}$ are the light speed and carrier frequency, respectively.
There are $N_{\rm c} = 3$ random scattering clusters ranging from $-\pi/2$ to $\pi /2$, each of which contains $N_{\rm s}=20$ sup-paths.
More details about the channel generation setting can be found in \cite{8590736}.
The antenna number of the BS $N_{\rm t}$ is set as 32 or 64.

We totally generate 1,000,000 CSI samples using MATLAB.
This dataset is randomly divided into training, validation, and test datasets, with 80\%, 10\%, and 10\% CSI samples, respectively.
In the simulation of the multi-cell system,
half of the dataset is chosen as  as the desired channel $\mathbf{h}$ and the other is the interfering channel $\mathbf{g}$. The cell number $K$ is set as ten.
In our simulation, we assume the desired and interfering channels obey the same distribution, which is similar to \cite{5613944} that models all the channels by the Rayleigh fading model, where each entry is a zero-mean unit-variance complex Gaussian independent and identically
distributed (i.i.d.) random variable according to $\mathcal{N}_{c}(0,1)$.
The interference source will be farther away, thereby making the interference signal very weak.
In our work, we use $\alpha_k \in [0,1]$ represents the strength ratio between the interfering and the desired signal. $\alpha_k = 0 $ means that there is no interference while $\alpha_k = 1$ means that the interfering and the desired signal have the same power.
Through adjusting the parameter $\alpha_k$, we can change the strength of the interference source.
In other words, though the desired and interfering channels obey the same distribution, the difference between them has been considered.

\subsubsection{NN training details}
All DL-based simulations are conducted using TensorLayer 1.11.0 \footnote{\url{https://tensorlayer.readthedocs.io/en/1.11.0/}} with one NVIDIA DGX-1 workstation.
The batchsize is set as 1,000 and the training epoch is 1,000.
The initial learning rate is 0.001 and it will decay by half when the loss does not decrease over 40 epochs.
The optimizer used in the simulation is the adaptive moment estimation optimizer (Adam).

At the inference period, the encoder at the user compresses the CSI using NNs and then quantizes the measurement vector.
However, once the BS receives the feedback bits, there is no need to infer the BF vector using the NN-based decoder.
The decoder creates a look-up table by recording the outputs of all possible codewords.
The BS can select the BF vector in the lookup table using the feedback codeword.

The DL-based algorithms are greatly dependent on the data.
If the DL-based algorithms, e.g., CsiFBnet, are deployed to the practical systems, the NNs should be offline-trained using the simulation data first.
Then, once NNs are embedded to the systems, a few training samples should be collected to online finetune the NNs.
When the communication business is idle, e.g., early morning, the BS collects the channel data and then updates the NN model.
More details about the practical deployment of the DL-based communication algorithms can be found in our previous work \cite{jiang2018artificial}.

\subsection{Performance of the CsiFBnet-s in the Single-cell System}
In this part, we evaluate the proposed CsiFBnet framework in the single-cell system.
We compare the proposed CsiFBnet-s with three different baseline algorithms.
Basline-1 denotes the method, which uses NNs to realize the feedback and adopts the traditional method to realize BF design, respectively.
Baseline-2 means the method, which adopts CS-based feedback and the traditional BF design strategy.
And Baseline-3 represents the method, which uses NNs to realize feedback and BF design, respectively.
The NN architecture used for baseline feedback algorithm is the same as that of CsiFBnet-s in Table \ref{NN1}, except that the last FC layer in the baseline is more complicated and with $2N_{\rm t}$ neurons, and the Lambda layer is removed.
The BF design in the Baselines-1,2 adopts the algorithm in \cite{7389996}, whose asymptotic computational complexity is $\mathcal{O}(N_{\rm t}^3)$.
The NNs for BF design in Baseline-3 are the same as those of decoder in the CsiFBnet-s.
During the training of Baseline-3, the NN input of the BF design is the imperfect CSI and the channel in the loss function is perfect.
Since the baseline methods have more complicated NNs and an extra traditional BF design algorithm, the proposed CsiFBnet-s has less computational complexity than the baselines.

\begin{figure}[t]
    \centering 
     \includegraphics[scale= 0.65]{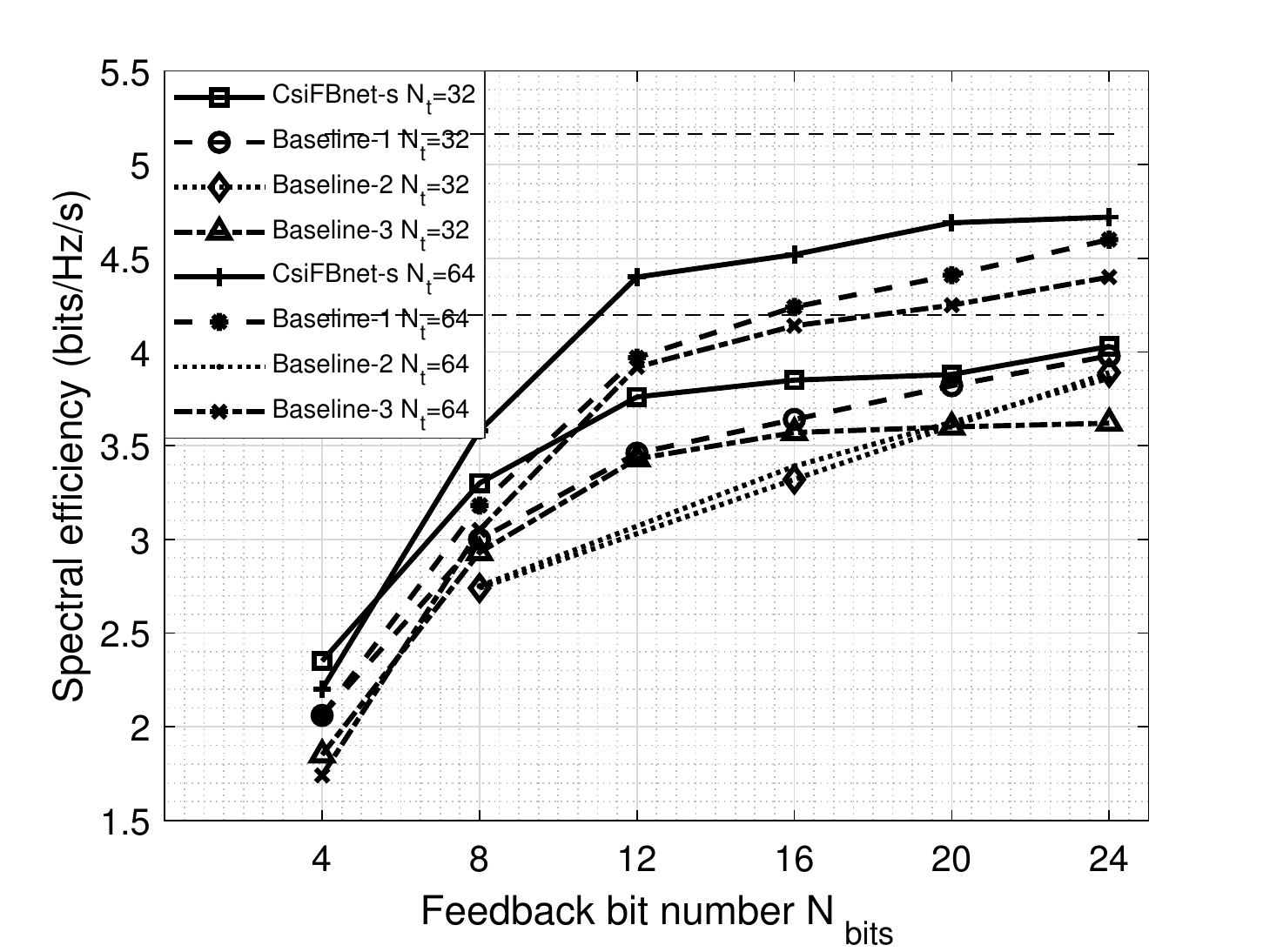}    
	\caption{\label{singleSE}Performance comparison of spectral efficiency between the CsiFBnet-s and baseline algorithms with $SNR=10{\rm dB}$.}
\end{figure}

  \begin{figure}[t]
    \centering 
     \includegraphics[scale=0.65]{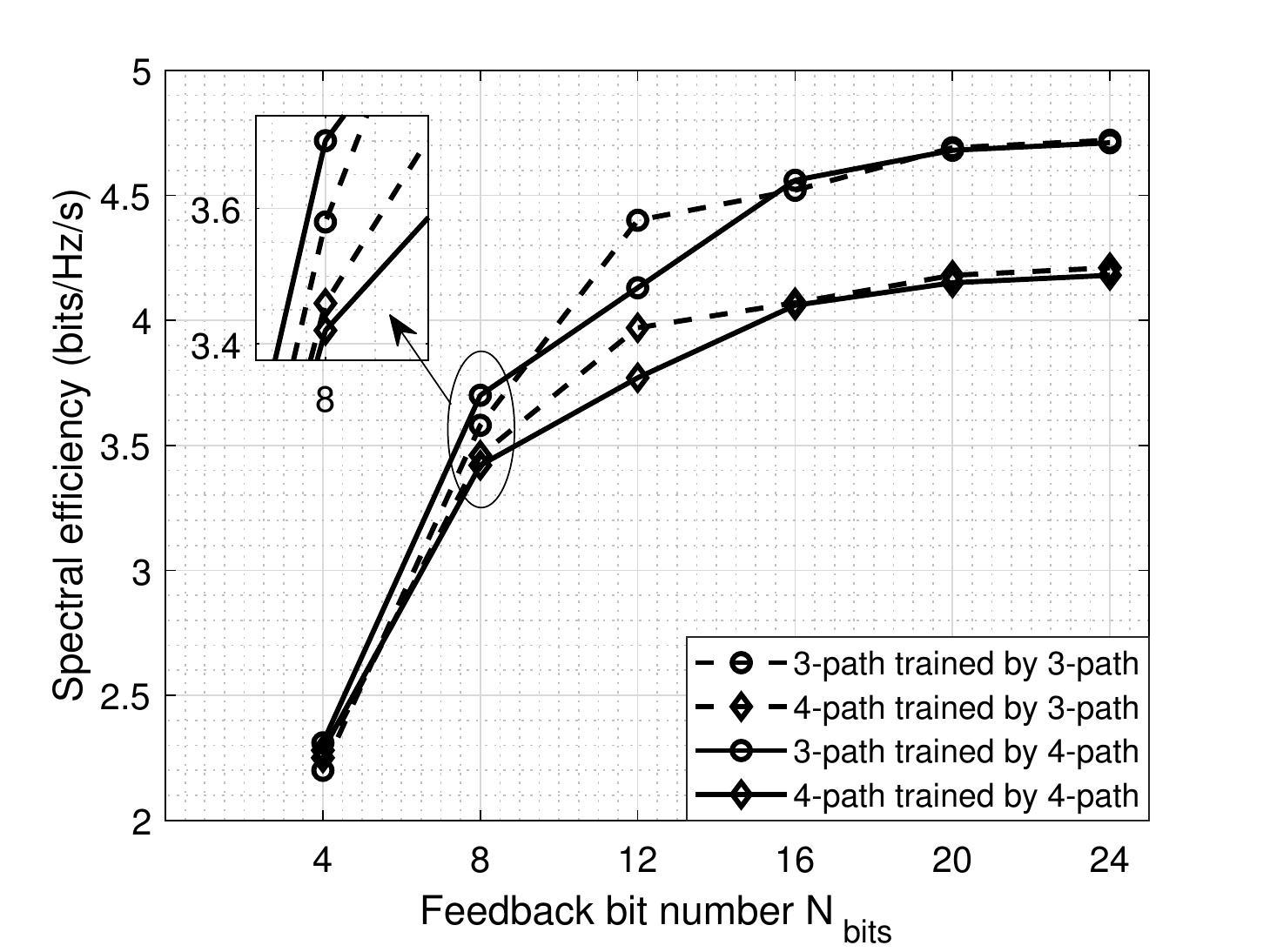}    
	\caption{\label{singleMismatch}Performance of the proposed CsiFBnet-s when there is a mismatch between the training and test datasets. `$i$-path trained by $j$-path' represents that the NNs are trained by the dataset with $j$ random scattering clusters and tested by the one with $i$ random scattering clusters.  }
\end{figure} 

Fig. \ref{singleSE} plots the spectral efficiency versus feedback-bit number $N_{\rm bits}$ of the proposed CsiFBnet-s and baseline algorithms, with $N_{\rm t} = 32,64$.
Baseline-1 performs better than Baseline-3, which is different from the simulation result in \cite{8847377}.
In that paper, the estimated channel quality is extremely low and the gap between the conventional and the DL-based algorithms becomes smaller with the increase of channel quality.
However, the BF design algorithm in \cite{7389996} requires perfect CSI.
Therefore, the DL-based algorithm outperforms the conventional one in \cite{8847377}.
For example, in that paper, when the pilot-to-noise power ratio is 0 dB, the NMSE of the estimated CSI is about 0.594 dB.
In our paper, even if the feedback bit number is only 8, the NMSE of the reconstructed CSI at the BS is -0.937 dB, which is better than that in \cite{8847377}.
Thus, Baseline-1 performs better than Baseline-3 in this paper.
The proposed framework always outperforms the baseline algorithms.
When the feedback overhead is greatly constrained, the performance gap is small.
For example, when the bit number is four and the BS antenna number $N_{\rm t}$ is 64, the proposed CsiFBnet-s only has a 0.14 bits/Hz/s performance improvement compared with Baseline-1.
From the perspective of the codebook-based feedback, there are only $2^4=16$ possible BF vectors for choice.
Therefore, no matter what algorithm is adopted, there is a very low  performance upper bound.
When the feedback-bit number is four, from (\ref{crEq}), the output of the encoder, i.e., measurement vector, only has one element and the compression ratio $\beta$ is 128.
From the perspective of information theory, though the proposed CsiFBnet forces the encoder extract the information that is significant for the BF design, the vector only with one element cannot contain enough information for the BF design, thereby making its performance outperform the baselines just a little.

However, when the feedback-bit number increases, the performance gap becomes larger.
The feedback NNs in the baseline do not know which information is useful for the BF design and just try to feed back all information, including the one useless for the BF design, thereby making its spectral efficiency performance worse than that of CsiFBnet-s by a large margin.
When the feedback-bit number is large, e.g., $N_{\rm bits}=20, 24$, the performance gap becomes smaller, which is due to that the feedback bits are enough for the NNs in the baselines to convert all information in the CSI.

It is widely accepted that the spectral efficiency will increase when the BS is equipped with more antenna.
However, when the feedback-bit number is small, e.g., $N_{\rm bits}=4$, the spectral efficiency of the system with $N_{\rm t} = 64$ is lower than that with $N_{\rm t} = 32$.
The CSI dimension of the former one is double of that of the latter one.
The CSI of the system with more antennas has more information needed to be fed back.
Therefore, when there are only four feedback bits, the system with $N_{\rm t} = 64$ antennas performs worse, which thereby shows the significance of the feedback design problem in the massive MIMO system.

We have added the upper bounds \footnote{The perfect CSI is assumed to be available for BF design at the BS.} for different antenna settings in the single-cell system.
The upper bounds with $N_{\rm t} = 32, 64$ are 4.2 bits/Hz/s and 5.18 bits/Hz/s , respectively.
When the transmitter antenna number is 32 and feedback bit number $N_{\rm bits}$ is 24, the gap between the upper bound and the system performance is very small.
However, when the BS is equipped with 64 antennas, the gap is still very large, which is because the feedback information increases with the dimension of the downlink channel.

\begin{figure}[t]
    \centering 
     \includegraphics[scale= 0.65]{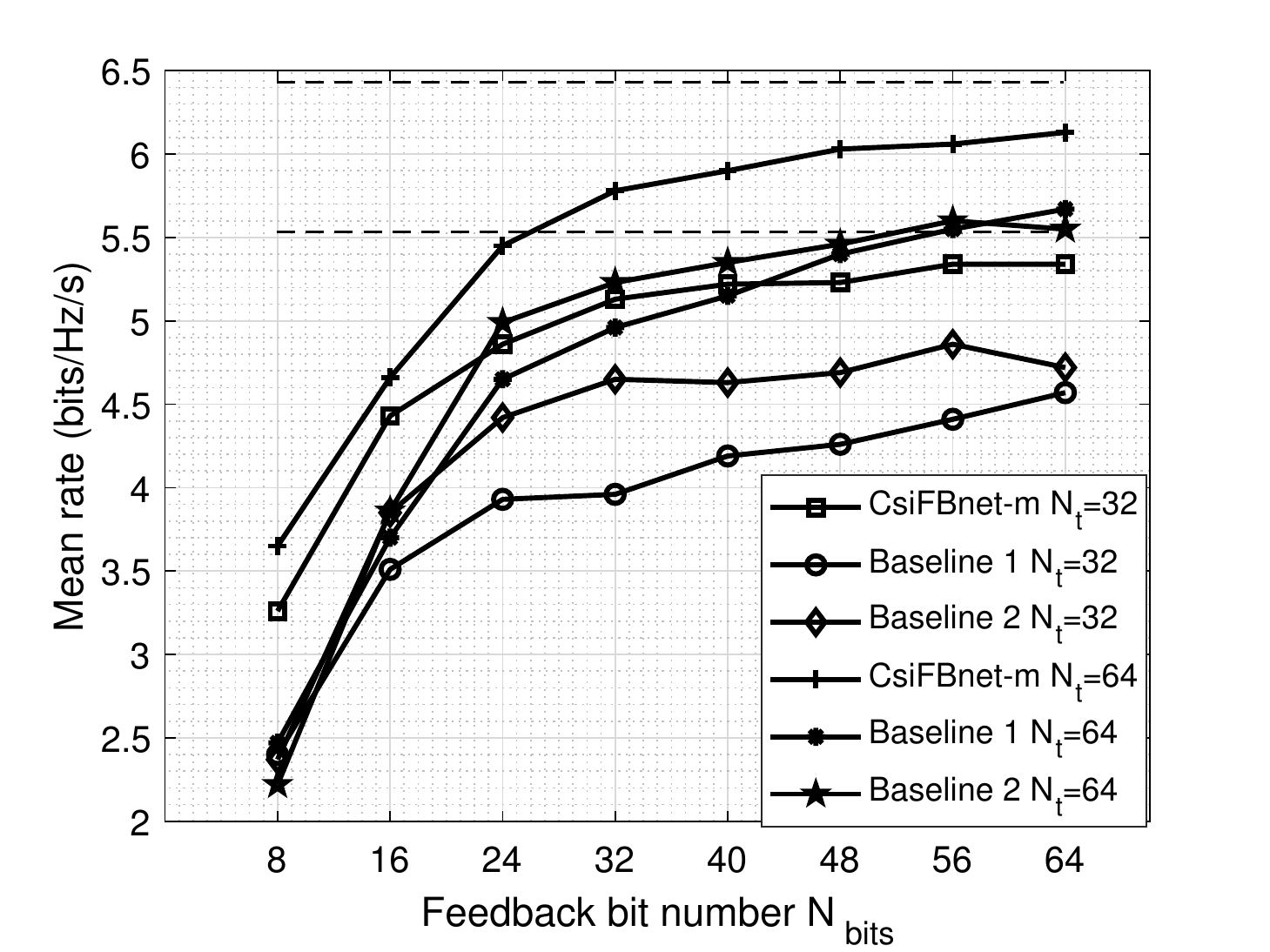}    
	\caption{\label{mcSE}Mean rate comparison among the proposed CsiFBnet-m and baseline algorithms with $\alpha = 0.1$, $\gamma_{\rm h} = \gamma_{\rm g}$, and  $SNR=15dB$. The two dashed lines represent the mean rates when the CSI is assumed to be perfectly fed back.}
\end{figure}

Since the proposed approach is based on a learning approach, one might expect a high sensitivity to the choice of the training dataset. 
Therefore, we introduce some artificial mismatch between the training and the test phases to test the robustness of the CsiFBnet-s.
Fig. \ref{singleMismatch} illustrates the spectral efficiency of the proposed CsiFBnet-s when there is a mismatch on the channel path number between the training and test datasets.
In the figure, `$i$-path trained by $j$-path' represents that the NNs are trained by the dataset with $j$ random scattering clusters and tested by the one with $i$ random scattering clusters.
It is widely accepted that the spectral efficiency of the user will increase if there are more paths.
However, in Fig. \ref{singleMismatch}, the one with 3 paths outperforms that with 4 paths, which is due to the CSI feedback.
More feedback bits are needed if the channel is more complicated, e.g., more paths.
In the simulation, we consider the scenario that the feedback bits are extremely limited, which may make feedback not accurate enough for the channel with more paths.
Therefore, in the imperfect feedback case, the user with fewer paths may perform better.
Then, from the figure, the mismatch between the training and test datasets has few effects on the performance.

We think that the NNs trained by 3-path feed back the information of three main paths while the NNs trained by 4-path feed back the information of four main paths.
And the latter NNs need more feedback bits due to the complexity of the channel.
When the feedback bits are not enough, the NNs, which only feed back the information of three main paths, may perform better.
Therefore, when the feedback bit number $N_{\rm bits}$ is 12, the NNs trained by 3-path channels outperform those trained by 4-path channels.

\subsection{Performance of the CsiFBnet-m in the Multi-cell System}

\subsubsection{Performance comparison between the CsiFBnet-m and baselines }

In this part, we give the mean rate performance and complexity comparison among the proposed CsiFBnet-m and two base algorithms.
In both of the two baseline algorithms, the feedback and the BF design are all realized by NNs.
The difference of them is that, during the training process, the input of the NNs for the BF design in the first algorithm is the perfect CSI while that in the second algorithm is the imperfect CSI, which is the output of the NNs for the feedback.
The NNs for CSI feedback used in the baseline algorithms are the same as these used in the single-cell scenario.
The NNs for the BF design are the same as these at the decoder of the CsiFBnet-m.
The loss function of the feedback process is MSE and the one of BF design is the same as that of the CsiFBnet-m.
The NN weights of the feedback module in the two baselines are the same but those of the BF design module are different since the training datasets are different in whether the feedback errors are considered.
The complexity comparison between the proposed CsiFBnet-m and the baselines is shown in Table \ref{complexity2}.
It is obvious that the parameter number and FLOPs are much smaller than these of the baselines.

\begin{table*}[t]
\caption{\label{complexity2}NN complexity comparison between the proposed CsiFBnet-m and the baselines.}
\centering
\resizebox{\textwidth}{!}{
\begin{tabular}{c|cc}
\hline \hline
            & Parameter number                                                                                                & FLOPs                                                                                                            \\ \hline
CsiFBnet-m & $(26 + 12/\beta_{\rm h} + 12/\beta_{\rm g}){N_{\rm t}^2} + (15 + 2/\beta_{\rm h} + 2/\beta_{\rm g}){N_{\rm t}}$ & $(52 + 24/\beta_{\rm h} + 24/\beta_{\rm g}){N_{\rm t}^2} - (15 + 2/\beta_{\rm h} + 2/\beta_{\rm g}){N_{\rm t}}$  \\ 
Baselines   & $(40 + 20/\beta_{\rm h} + 20/\beta_{\rm g}){N_{\rm t}^2} + (31 + 2/\beta_{\rm h} + 2/\beta_{\rm g}){N_{\rm t}}$ & $(100 + 40/\beta_{\rm h} + 40/\beta_{\rm g}){N_{\rm t}^2} - (31 + 2/\beta_{\rm h} + 2/\beta_{\rm g}){N_{\rm t}}$ \\ \hline \hline
\end{tabular}
}
\end{table*}

\begin{figure}[t]
    \centering 
     \includegraphics[scale= 0.65]{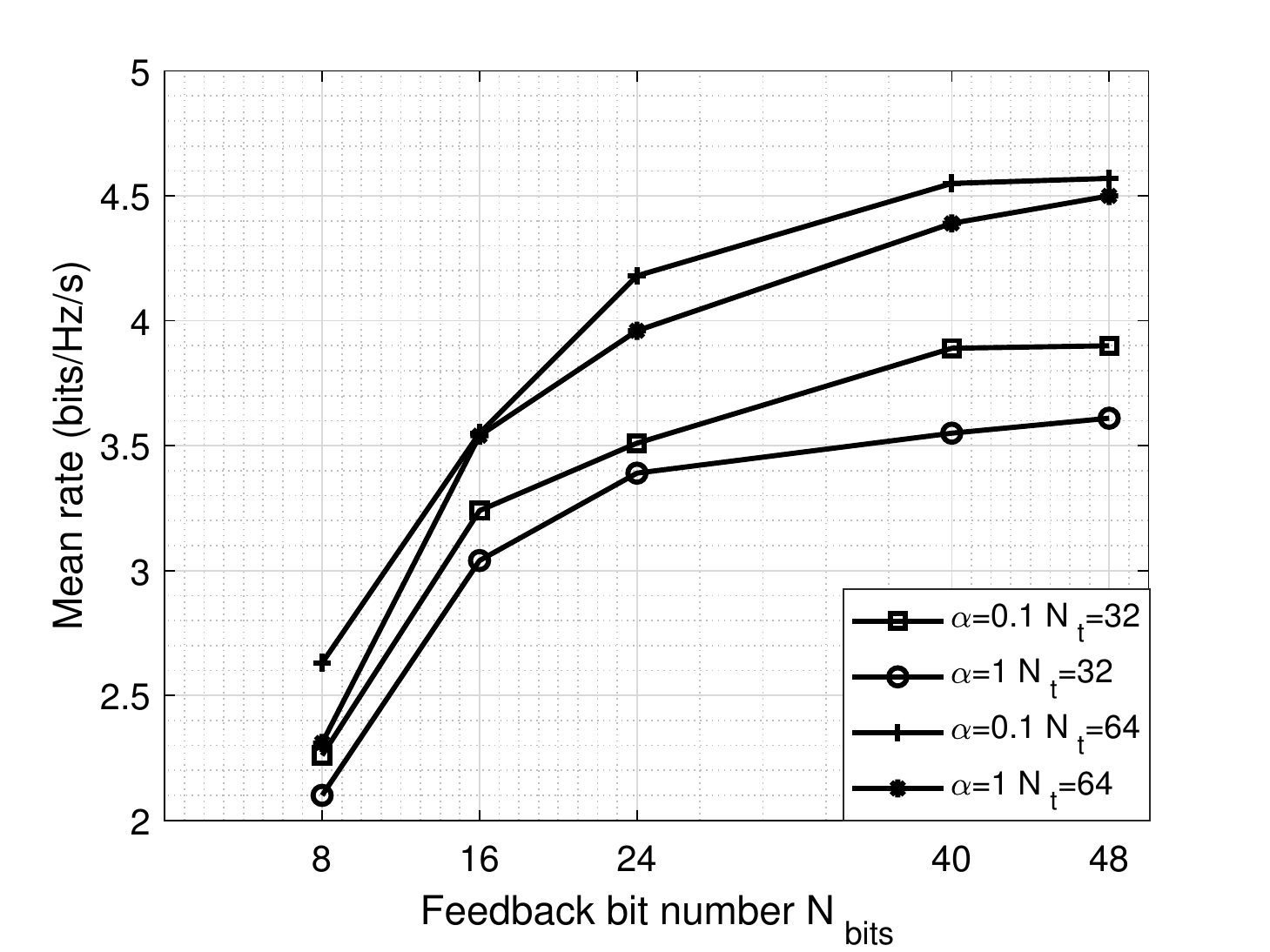}    
	\caption{\label{alpha}Performance of the proposed CsiFBnet-m with different $\alpha$ where $\gamma_{\rm h} = \gamma_{\rm g}$, and  $SNR=10dB$.}
\end{figure}

In the simulations, we set $\alpha$ and $SNR$ to be 0.1 and 15 dB, respectively, and assume that the desired and the interfering channel occupy the equal feedback overhead, i.e., $\gamma_{\rm h} = \gamma_{\rm g}$.
Fig. \ref{mcSE} plots the mean rate versus feedback-bit number for three algorithms.
The second baseline algorithm outperforms the first one by a margin, which shows the importance of taking imperfect CSI into consideration during the BF design process.
However, when the feedback-bit number is large, e.g., $N_{\rm bits}=64$, their rates are similar because enough bits are fed back to the BS for reconstruction, that is, the reconstructed at the BS is accurate enough.

Different from the single-cell system, where the performance gap is smaller with very small or large $N_{\rm bits}$, the CsiFBnet-m outperforms the baselines by a large margin with different $N_{\rm bits}$, which shows the great potential of the CsiFBnet framework in the multi-cell system.
We also plot the upper bounds in Fig. \ref{mcSE}.
The two dashed lines represent the mean rates when the CSI is assumed to be perfectly fed back, which can be regarded as the upper bounds of the multi-cell system.
When the antenna number $N_{\rm t}$ is 32 and 64, the rate upper bounds are 5.55 bits/Hz/s and 6.46 bits/Hz/s, respectively.
When the feedback-bit number $N_{\rm bits}$ is above 48, there is less performance improvement with the increase of the bit number and the rates achieved by CsiFBnet-m with $N_{\rm bits}=32,64$  are 5.34 bits/Hz/s and 6.13 bits/Hz/s, respectively, which are lower than the upper bounds by 0.21 bits/Hz/s and 0.33 bits/Hz/s, respectively.

Fig. \ref{alpha} shows the effect of the channel strength ratio $\alpha$ on the rate performance.
$\alpha$ represents the strength relation of desired and the interfering signals the user receives from its corresponding cell and the nearby cell, respectively. 
From the figure, as imagined, we can see that the system with low $\alpha$ outperforms that with high $\alpha$, which means that the cell with high transmit power will have more interference on the user at the adjacent cell.

\subsubsection{Effect of the feedback-bit allocation on the mean rate  }

\begin{figure}[t]
    \centering 
     \includegraphics[scale= 0.65]{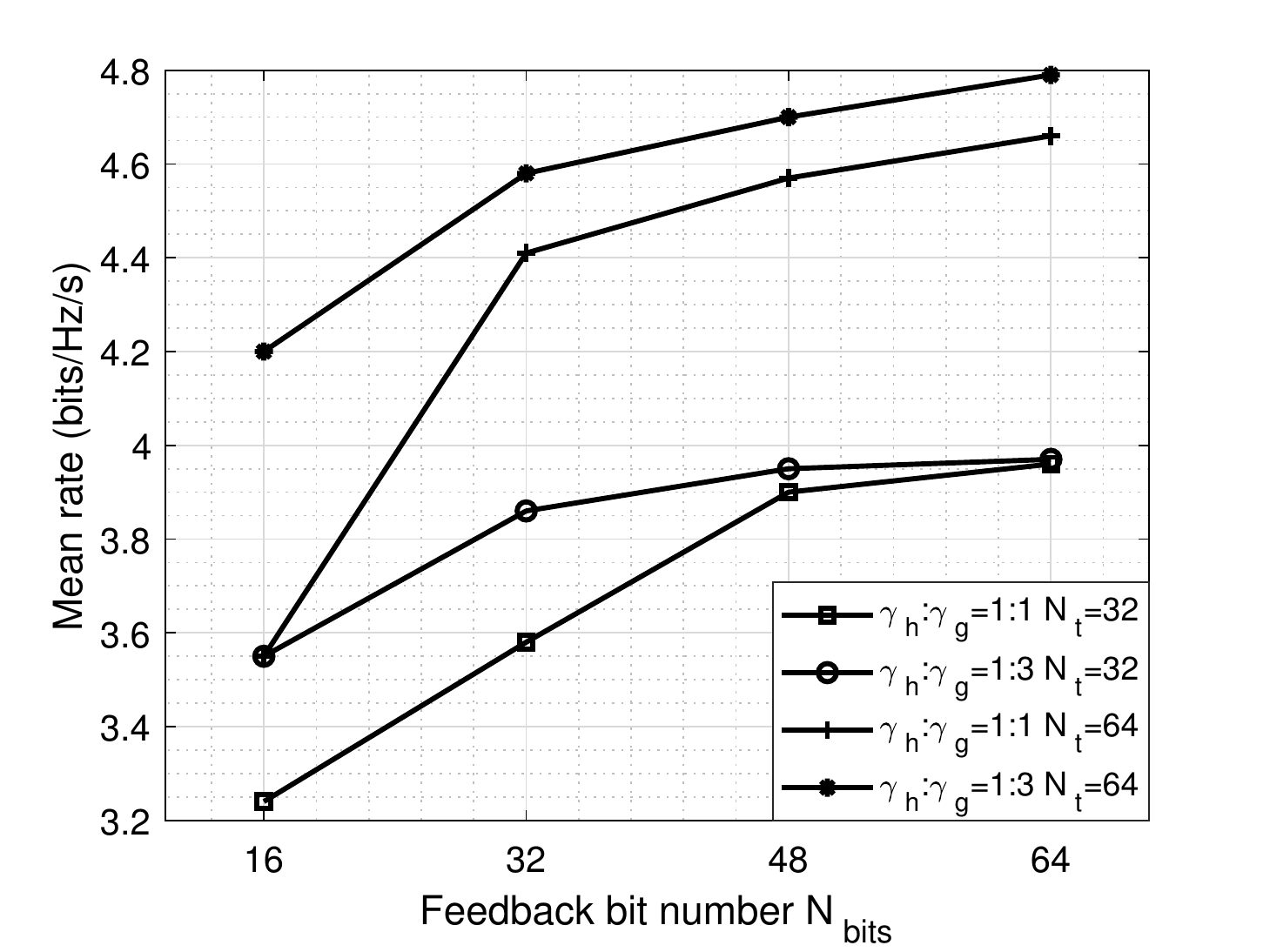}    
	\caption{\label{di}Performance of the proposed CsiFBnet-m with different $\gamma_{\rm h}:\gamma_{\rm g}$ where $\alpha=0.1$, and  $SNR=10dB$.}
\end{figure}

As shown in Fig. \ref{multi-cell-exchange}, there are two kind of CSI, i.e., $\mathbf{h}_k$ and $\mathbf{g}_{k+1}$, needed to be fed back for the $k^{\rm th}$ user, which causes a feedback-bit allocation problem and has been well studied in the existing works, e.g., \cite{5613944}.
Fig. \ref{di} shows the rate performance of the CsiFBnet-m with different  feedback-bit allocation strategies.
As defined in (\ref{crEq}), $\gamma_{\rm h}:\gamma_{\rm g}=1:3$ represents that the feedback-bit number of $\mathbf{h}_{k}$ is three times that of $\mathbf{g}_{k+1}$ and $\gamma_{\rm h}:\gamma_{\rm g}=1:1$ denotes that they occupy the same feedback overhead.
We here set $\alpha$ and $SNR$ as 0.1 and 10 dB, respectively.
Since the signal strength of the desired channel is ten times that of the interfering channel, an intuitive idea is that more feedback bits should be allocated for the feedback of $\mathbf{h}_{k}$.
It can be seen in Fig. \ref{di} that the rate performance with $\gamma_{\rm h}:\gamma_{\rm g}=1:3$ outperforms that with $\gamma_{\rm h}:\gamma_{\rm g}=1:1$.
However, with the increase of the feedback-bit number, the performance gap becomes smaller.
For example, when the BS antenna number $N_{\rm t}$ is 32 and the total feedback-bit number $N_{\rm bits}$ is 64, there are two bit allocation strategies: 1) 32 bits for $\mathbf{h}_k$ and 32 bits for $\mathbf{g}_{k+1}$, and 2) 48 bits for $\mathbf{h}_k$ and 16 bits for $\mathbf{g}_{k+1}$.
From the figure, the performance gap between two feedback strategies is negligible.
The reason is similar to that in Fig. \ref{singleSE}, where the feedback bits are enough for the BF design and so the allocation strategy has less effect on the rate performance.
When the feedback-bit number is very small, the performance gap is large since the desired channel has more effects on the user.

\begin{figure}[t]
    \centering 
     \subfigure[Robustness of the CsiFBnet-m to the SNR.]{\label{sf1}
	\includegraphics[scale=0.65]{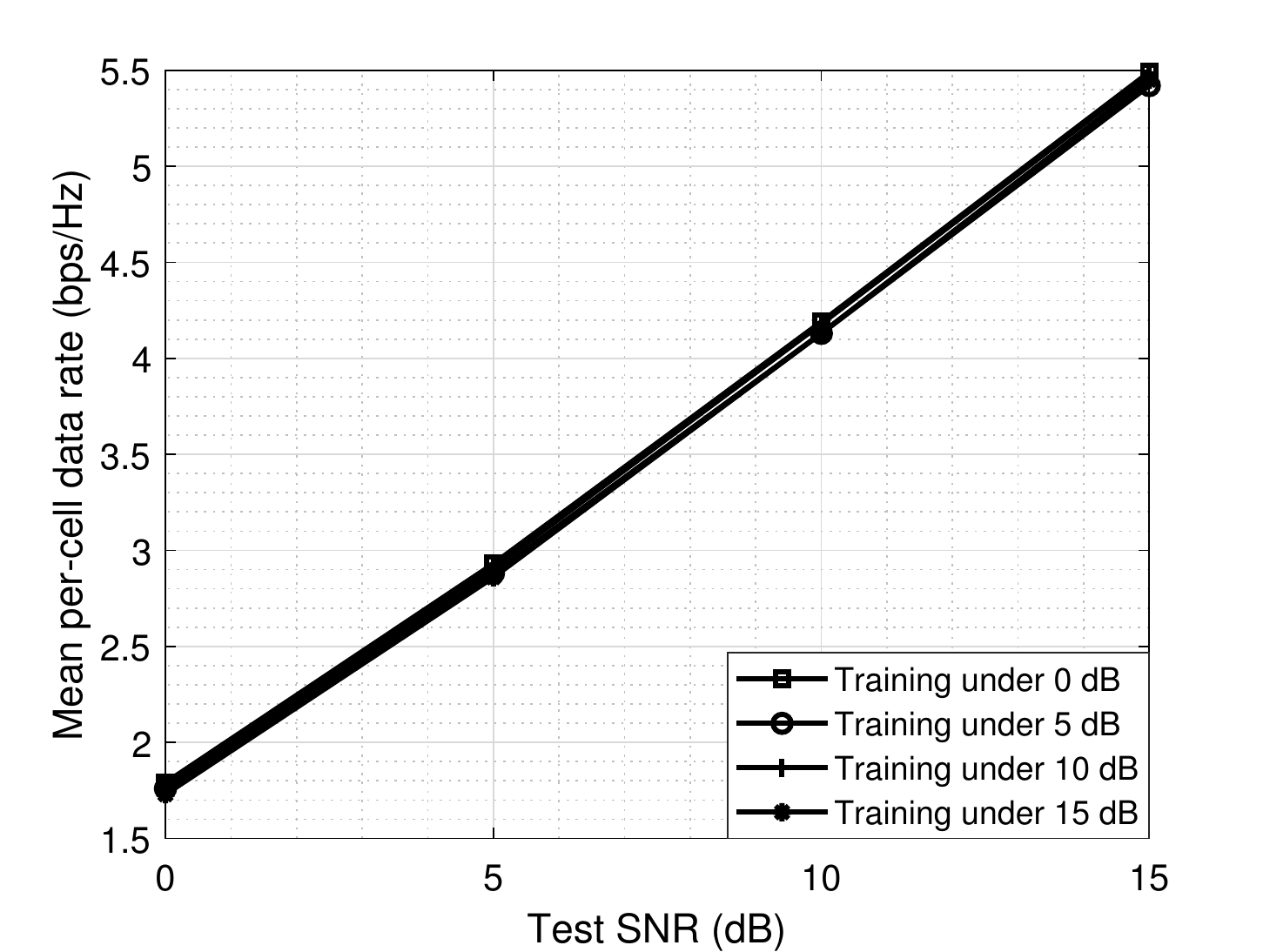}	}
\subfigure[Robustness of the CsiFBnet-m to the $\alpha$.]{\label{sf2}
	\includegraphics[scale=0.65]{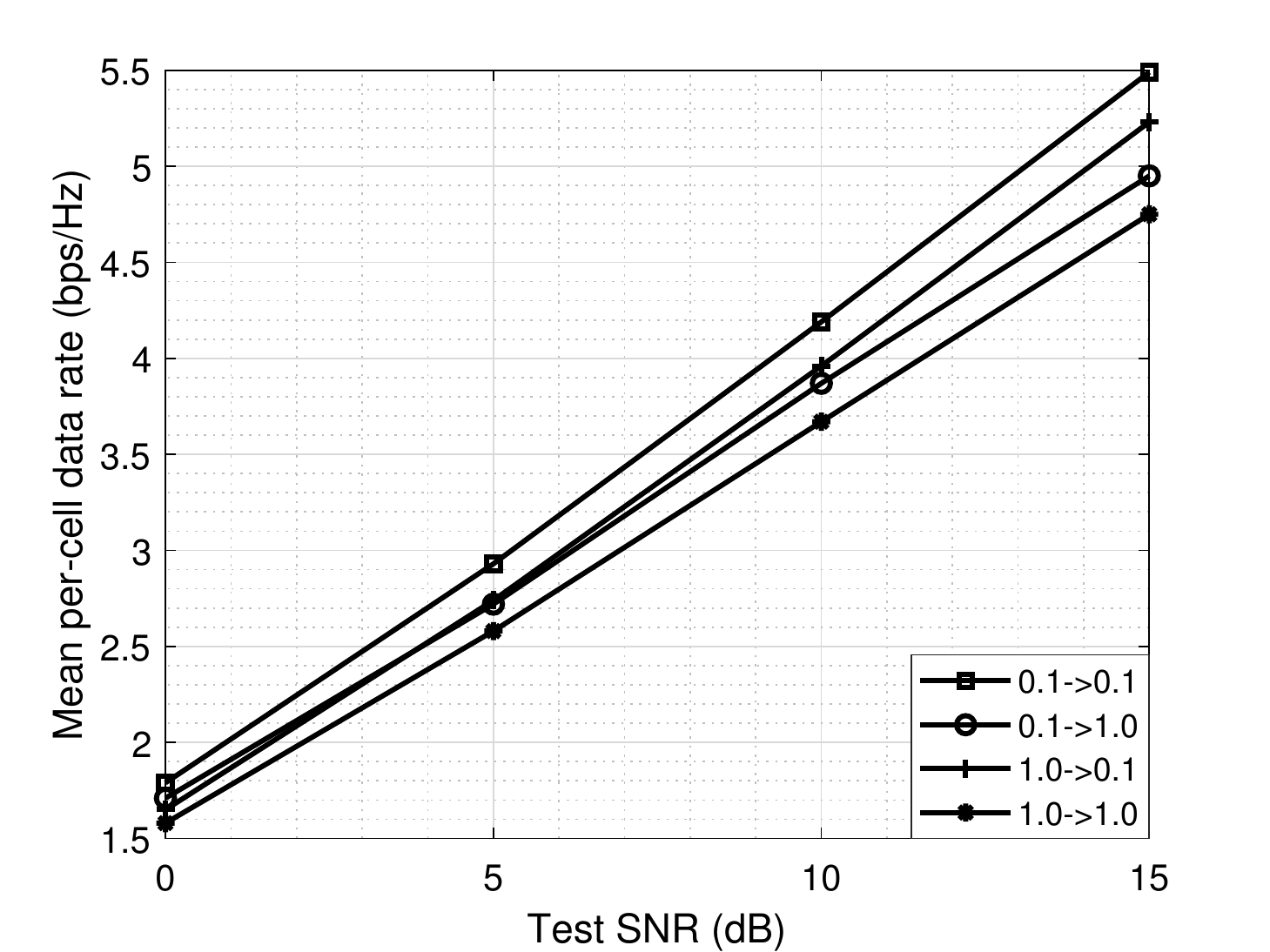}	}
	\caption{\label{robustness} 
	Robustness of the proposed CsiFBnet-m. The BS is equipped with $N_{\rm t} = 64$ transmit antennas and the feedback-bit number $N_{\rm bits}$ is 24. In the first subfigure, $\alpha$ is set as 0.1. In the second subfigure, the training SNR is 0 dB. 0.1->1.0 means that the training and the test $\alpha$ are 0.1 and 1.0, respectively.} 
\end{figure} 

\subsubsection{Robustness of the proposed CsiFBnet-m}
We first evaluate the mismatch problem of the training and the test SNRs in the Fig. \ref{robustness}.
The first subfigure plots the mean rate versus test SNR for the NNs trained under different SNRs.
In this simulation, the BS is equipped with 64 antennas, the 
feedback-bit number is 24, and the desired and the interfering channels occupy equal feedback overhead.
From the subfigure, the rate performance under all test SNRs is similar no matter what the training SNR is, which shows the great robustness of the proposed CsiFBnet framework.
During the simulation, we set the training SNR to be $+\infty$, that is, $\frac{1}{\rho_{k},(\mathrm{d})} = 0$ in (\ref{snrEq}).
Unexpectedly, the data rate performance is very poor and the training loss can not converge.

Fig. \ref{sf2} shows the robustness of the CsiFBnet-m to $\alpha$.
In this simulation, the antenna number is $N_{\rm t}=64$, the training SNR is 0 dB, the feedback-bit number is $N_{\rm bits}=24$, and $\alpha$ is set to be 0.1 or 1.0.
Different from the SNR robustness, the mean rate is sentive to $\alpha$.
Even if the test $\alpha$ is 1.0, the system trained with $\alpha=0.1$ still outperforms that trained with $\alpha=1.0$.
Therefore, during the practical deployment, the training $\alpha$ should be selected through simulations.

\section{Conclusion}
\label{s6}
In this paper, we propose a DL-based CSI feedback framework for BF design.
The key idea of the proposed CsiFBnet is to maximize the performance gain as the optimization of the feedback instead of the feedback accuracy. 

First, we consider a single-cell system.
The encoder followed by a uniform quantizer at the user compresses the CSI and the decoder at the BS generates the BF design from the quantized measurement vector.
Then, we extend the proposed CsiFBnet framework to the soft hand-off multi-cell system, where there are the desired and the interfering channels .
Correspondingly, there are two encoder modules at the users to compress these two kind CSI, respectively.
The decoder at the BS generates the BF vector using the information from the two measurement vectors.
Simulation results show the performance improvement achieved by the proposed CsiFBnet framework.
In the multi-cell system, the feedback-bit allocation has great effects on the system performance.
Besides, the CsiFBnet-m is robust to SNR but senstive to $\alpha$, which gives a guideline for future research.

\ifCLASSOPTIONcaptionsoff
  \newpage
\fi

\bibliographystyle{IEEEtran}
\bibliography{CsiFBnet}

\end{document}